\documentclass[preprint]{aastex63}
\shortauthors{bora et al.}
\usepackage[utf8]{inputenc} 
\usepackage[T1]{fontenc} 
\usepackage{amsmath,amssymb}

\usepackage{graphicx}
\usepackage{appendix}
\begin{document}

\title{Comparison of the Hall Magnetohydrodynamics and Magnetohydrodynamics evolution of a flaring solar active region}
\author{K. Bora} 
\affil{Udaipur Solar Observatory, Physical Research Laboratory, Dewali, Bari Road, Udaipur-313001, India}
\affil{Discipline of Physics, Indian Institute of Technology, Gandhinagar-382355, India}
\author{R. Bhattacharyya}
\affil{Udaipur Solar Observatory, Physical Research Laboratory, Dewali, Bari Road, Udaipur-313001, India}
\author{Avijeet Prasad}
\affil{Center for Space Plasma \& Aeronomic Research, The University of Alabama in Huntsville, Huntsville, AL 35899, USA}
\affil{Institute of Theoretical Astrophysics, University of Oslo, Postboks 1029, Blindern No-0315 Oslo, Norway}
\affil{Rosseland Centre for Solar Physics, University of Oslo, PO Box 1029, Blindern 0315, Oslo, Norway}
\author{Bhuwan Joshi}
\affil{Udaipur Solar Observatory, Physical Research Laboratory, Dewali, Bari Road, Udaipur-313001, India}
\author{Qiang Hu}
\affil{Center for Space Plasma \& Aeronomic Research, The University of Alabama in Huntsville, Huntsville, AL 35899, USA}
\affil{Department of Space Science, The University of Alabama in Huntsville, Huntsville, AL 35899, USA}

\accepted{on November, 20 2021}

\begin{abstract}
This work analyzes the
Hall magnetohydrodynamics (HMHD) and magnetohydrodynamics (MHD)
numerical simulations of a flaring solar active region as a 
testbed while idealizing the coronal Alfv\'en speed
to be of two orders of magnitude lesser. HMHD supports faster magnetic reconnection and shows richer
complexity in magnetic field line evolution compared to the MHD. The magnetic reconnections triggering the flare
are explored by numerical simulations augmented with relevant
multi-wavelength observations. The initial coronal magnetic field
is constructed by non-force-free extrapolation of photospheric
vector magnetic field.  Magnetic structure involved in the flare
is identified to be a flux rope, with its overlying magnetic field
lines constituting the quasi-separatrix layers (QSLs) along with a
three-dimensional null point and a null line. Compared to the MHD
simulation, the HMHD simulation shows a higher and faster ascend
of the rope together with the overlying field lines, which further
reconnect at the QSL located higher up in the corona. The foot
points of the field lines match better with the observations for the
HMHD case with the central part of the flare ribbon located at the
chromosphere. Additionally, field lines are found to rotate in a
circular pattern in the HMHD, whereas no such rotation is seen
in the MHD results. Interestingly, plasma is also observed to be
rotating in a co-spatial chromospheric region, which makes the HMHD
simulation more credible. Based on the aforementioned agreements,
HMHD simulation is found to agree better with observations and,
thus, opens up a novel avenue to explore.  \end{abstract}


\section{Introduction}
\label{intro}
The astrophysical plasmas characterized by high Lundquist number
$S\equiv Lv_A/\eta$ ($L\equiv$ length scale of the magnetic field
\textbf{B} variability, $v_A\equiv$ Alfv\'en speed, and $\eta\equiv$
magnetic diffusivity) satisfy the Alfv\'en's flux-freezing theorem in presence of laminar plasma flow, ensuring  magnetic field lines to be tied to  fluid
parcels \citep{Alfven}. The scenario is different 
in a turbulent magnetofluid, see \citet{Vishnaic1999, Vishnaic2000, Eyink} 
for details.
An inherent large $L$ implies large $S$ and
ensures the flux freezing in the astrophysical plasmas. Particularly,
the solar corona with global $L\approx 100 ~\rm Mm$, $v_{A}\approx10^{6}$
ms$^{-1}$, B$\approx10$ G, and $\eta\approx1$ m$^2$s$^{-1}$ (calculated
using Spitzer resistivity) has a $S\approx10^{14}$ \citep{Aschwanden}.
However, the coronal plasma also exhibits diffusive behavior in the
form of solar transients---such as solar flares, coronal mass
ejections (CME), and jets. All of these are manifestations of
magnetic reconnections that in turn lead to dissipation of magnetic
energy into heat and kinetic energy of plasma flow, accompanied by
a rearrangement of magnetic field lines \citep{Arnab}. The magnetic
reconnections being dissipative processes, their onset is due to
the generation of small scales in consequence of large-scale dynamics,
ultimately increasing the magnetic field gradient and thereby
resulting in intermittently diffusive plasma. The small scales may
naturally occur as current sheets (CSs) \citep{ParkerECS},
magnetic nulls \citep{Parnell96,Ss2020} and quasi-separatrix layers 
(QSLs) \citep{Demoulin, avijeet2020}, or can develop spontaneously 
during the evolution of the magnetofluid. Such spontaneous developments
(owing to discontinuities in the magnetic field) are expected from 
Parker’s magnetostatic theorem \citep{ParkerECS} and have also been 
established numerically by MHD simulations \citep{Ss2020, DKRB,
SKRB, Sanjay2016, SK2017, avijeet2017, avijeet2018, Ss2019,
Sanjay2021}. Identification of the small (viz. the dissipation) 
scale depends on the specific physical system under consideration. 
For example, the length scale at which the reconnection occurs is 
found to be $L_{\eta}\equiv\sqrt{\tau_{d}\eta}\approx$32~m, based on 
$\eta\approx1$ m$^2$s$^{-1}$ and the magnetic diffusion time scale 
$\tau_{d}$ approximated by the impulsive rise time of hard X-ray 
flux $\approx 10^3$ s \citep{PF200} during a flare. Consequently, 
the estimated ion inertial length scale 
$\delta_i\approx 2.25$ m in the solar corona \citep{PF200} suggests 
that the order of the dissipation term, $1/S\approx 10^{-5}$ (approximated
with $L_{\eta}$), is 
 smaller than the order of the Hall term, $\delta_i/L_\eta\approx 10^{-2}$,
in standard dimensionless induction equation 
\citep{Westerberg07, 2021ApJ...906..102B}
\begin{equation}
\label{inducresist}
\frac{{\partial\bf{B}}}{\partial t} = 
\nabla\times \left({\bf{v}}\times{\bf{B}}\right)
-\frac{1}{S}\nabla\times{\bf{J}}
-\frac{\delta_i}{L_\eta}\nabla\times\left({\bf{J}}\times{\bf{B}}\right)~,
\end{equation}
where ${\bf{J}}(=\nabla\times{\bf{B}})$ and ${\bf{v}}$ are the 
volume current density and the plasma flow velocity, respectively.
This difference in the order of magnitude irrefutably indicates the
importance of the Hall term in the diffusive limit {\bf{ \citep{BIRN, BhattacharjeeReview}}} of the solar
coronal plasma which, further signifies that the HMHD can play a
crucial role for coronal transients as the magnetic reconnections
are their underlying mechanism. Importantly, the aforesaid activation 
of the Hall term only in the diffusive limit is crucial in setting up a
HMHD based numerical simulation, invoked latter in the paper.

Important insight into magnetic reconnection can be gained by casting 
(\ref{inducresist}) in absence of dissipation, as

\begin{equation}
 \label{inducresist1}
\frac{{\partial\bf{B}}}{\partial t} = 
\nabla\times \left({\bf{w}}\times{\bf{B}}\right)~,   
\end{equation}
\noindent following \citet{hornig-schindler}. The velocity ${\bf{w}}={\bf{v}}-\delta_i/L_\eta{\bf{J}}$,
which is also the electron fluid velocity, conserves magnetic flux \citep{schindler} and topology \citep{hornig-schindler} since 
field lines are tied to it. Consequently, field lines slip out from the fluid parcels advecting with velocity 
{\bf{v}} to which the lines are frozen in ideal MHD. Importantly, the resulting breakdown of the flux freezing is 
localized to the region where current density is large and the Hall term is effective. Because of the slippage, two fluid parcels do not remain connected with the same field lines over time---a change in field line connectivity. Quoting \citet{schindler}, such localized breakdown of flux freezing along with the resulting change in connectivity can be considered as the basis of reconnection \citep{axford}. Additional slippage of field lines
occur in presence of the dissipation term, but with a change in magnetic topology. The present paper extensively relies on this interpretation of reconnection as the slippage of magnetic field lines and the resulting change in magnetic connectivity. 

The importance of HMHD is by no means limited to
coronal transients. For example, HMHD is important in the Earth's
magnetosphere, particularly at the magnetopause and the magnetotail
where CSs are present \citep{Mozer2002}. Generally, the HMHD is
expected to support faster magnetic reconnections, yet without
directly affecting the dissipation rate of magnetic energy and
helicity by the Hall term in the induction equation \citep{PF200, chenshi}. The faster reconnection may be associated with a more effective 
slippage of field lines in HMHD compared to the resistive MHD, compatible 
with the arguments presented earlier. Nevertheless,
these unique properties of the HMHD are expected to bring
subtle changes in the dynamical evolution of plasma, particularly
in the small scales dominated by magnetic reconnections, presumably
bringing a change in the large scales as a consequence. Such subtle
changes were found in the recent HMHD simulation
\citep{2021ApJ...906..102B}, performed by extending the computational
model EULAG-MHD \citep{PiotrJCP} to include the Hall effects.
Notably, the faster reconnection compared to MHD led to a breakage
of a magnetic flux rope, generated from analytically constructed
initial bipolar magnetic field lines \citep{Sanjay2016}.  In turn,
the flux rope breakage resulted in the generation of magnetic
islands as theorized by \citet{Shibata}. Clearly, it is compelling 
to study the HMHD evolution in a more realistic scenario with the 
initial magnetic field obtained from a solar magnetogram. To attain
such an objective, we select the recently reported active 
region (AR) NOAA 12734 by \citet{2021Joshi} that produced a C1.3 class 
flare.

In absence of reliable direct measurement of the coronal magnetic field, 
several extrapolation models such as nonlinear force-free
field (NLFFF) \citep{2008Wglman, 2012WglmnSakurai} and non-force-free
field (non-FFF) \citep{HuDas08, Hu2010} have been developed to construct
the coronal magnetic field using photospheric magnetograms. The
standard is the NLFFF, and the recent data-based MHD simulations
initialized with it have been reasonably successful in simulating
the dynamics of various coronal transients \citep{2013Jiang,
2014NaturAm, 2014Innoue, 2016Savcheva}. However, the NLFFF
extrapolations require to treat the photosphere as force-free, while
it is actually not so \citep{Gary}. Hence, a ``preprocessing technique''
is usually employed to minimize the Lorentz force on the photosphere
in order to provide a boundary condition suitable for NLFFF
extrapolations \citep{2006SoPhWgl, 2014SoPhJiang} and thereby
compromising the reality. Recently, the non-force-free-field (non-FFF)
model, based on the principle of minimum energy dissipation rate
\citep{bhattaJan2004, bhattaJan2007}, has emerged as a plausible
alternative to the force-free models \citep{HuDas08, Hu2010,
2008ApJHu}. In the non-FFF model, the magnetic field \textbf{B} satisfies
the double-curl-Beltrami equation \citep{MahajanYoshida} and the
corresponding Lorentz force on the photosphere is non-zero while
it decreases to small values at the coronal heights \citep{avijeet2018,
Ss2019, avijeet2020}---concurring with the observations. In this
paper, we use non-FFF extrapolation \citep{Hu2010} to obtain the
magnetic field in corona using the photospheric vector magnetogram
obtained from the Helioseismic Magnetic Imager (HMI) \citep{HMI}
onboard the Solar Dynamics Observatory (SDO) \citep{SDO}.

The paper is organized as follows. Section \ref{obs} describes the
flaring event in AR NOAA 12734, section \ref{extrapolation} presents
magnetic field lines morphology of AR NOAA 12734 along with the
preferable  sites for magnetic reconnections such as QSLs, 3D null
point, and null-line found from the non-FFF extrapolation. Section
\ref{simulation-results} focuses on the numerical model, numerical
set-up and the evolution of magnetic field lines obtained from
the extrapolation along with their realizations in observations. 
Section \ref{summary} highlights the key findings.

\section{Salient features of the C1.3 class flare in AR NOAA 12734}
\label{obs}
\begin{figure}
\centering
\includegraphics[width=1\textwidth]{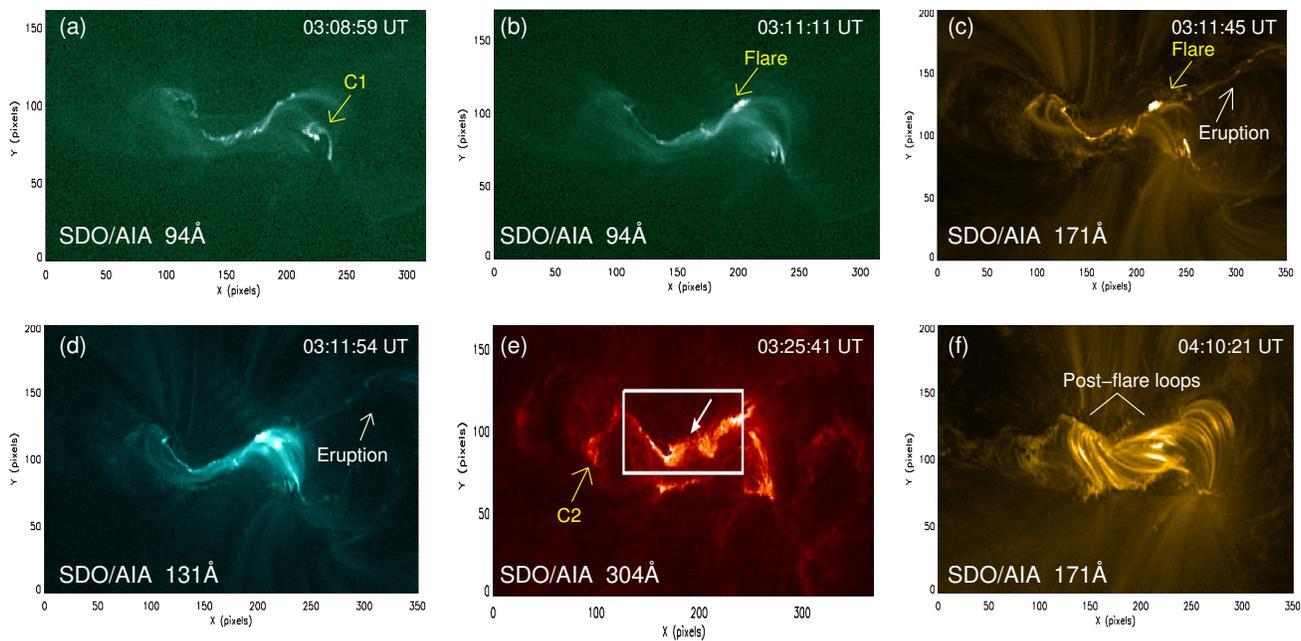}
\caption{Panels (a)-(f) are SDO/AIA images showing the multi-wavelength
observations of the flaring active region AR NOAA 12734. Panel (a)
shows the quasi circular brightening at the western part of AR prior
to the flare (marked by C1). Panels (b)-(d) show the initiation of
the flare followed by eruption (indicated by yellow arrow). Panel
(e) shows the circular structure after eruption at the eastern part
of AR (marked by C2) and the W-shaped flare ribbon (enclosed by
white box). Panel (f) shows the post-flare loops.}
\label{observations}
\end{figure}

The AR NOAA 12734 produced an extended C1.3 class flare
on March 08, 2019 \citep{2021Joshi}. The impulsive phase of the
flare started at 03:07 UT as reported in the Figure 3 of
\citet{2021Joshi}, which shows the X-ray flux in the 1-8 {\AA} and
0.5-4 {\AA} detected by the Geostationary Operational Environmental
Satellite (GOES) \citep{Gracia}. The flux evinces
two subsequent peaks after the onset of the flare,
one around 03:19 UT and another roughly around 03:38 UT.   \citet{2021Joshi} 
suggested the eruptive event to take place in a coronal sigmoid
with two distinct stages of energy release. Additional observations
using the multi-wavelength channels of Atmospheric Imaging Assembly
(AIA) \citep{AIA} onboard SDO are listed below to highlight important
features pertaining to simulations reported in this paper. Figure
\ref{observations} illustrates a spatio-temporal
observational overview of the event. Panel (a)
shows the remote semicircular brightening (C1) prior to the impulsive
phase of the flare (indicated by the yellow arrow). Panels (b) to (d)
indicate the flare by yellow arrow and the eruption by the white arrow
in the 94 {\AA}, 171 {\AA}, and 131 {\AA} channels respectively.
Notably, the W-shaped brightening appears in panels (b) to (d) along
with the flare in different wavelength channels of SDO/AIA. Panel
(e) shows the circular structure of the chromospheric material (C2)
during the impulsive phase of the flare.  It also highlights the
developed W-shaped flare ribbon (enclosed by the white box) which has 
a tip at the center (marked by the white arrow).  Panel (f) depicts
the post-flare loops in 171 {\AA} channel, indicating the post-flare
magnetic field line connectivity between various negative and
positive polarities on the photosphere.

\section{non-FFF Extrapolation of the AR NOAA 12734}
\label{extrapolation}
As stated upfront, the non-FFF extrapolation technique proposed by
\citet{HuDas08} and based on the minimum dissipation rate theory
(MDR) \citep{bhattaJan2004, bhattaJan2007} is used to obtain the
coronal magnetic field for the AR NOAA 12734. The extrapolation
essentially solves the equation
\begin{eqnarray}
\label{tc}
\nabla\times\nabla\times\nabla\times \textbf{B}+a_1 \nabla\times\nabla\times 
\textbf{B}+b_1 \nabla\times\textbf{B}=0~, 
\end{eqnarray}
where parameters $a_1$ and $b_1$ are constants. Following 
\citep{Hu2010}, the field is constructed as  
\begin{eqnarray}
\textbf{B}=\sum_{i=1,2,3} \textbf{B}_{i}~,~~ \nabla\times \textbf{B}_{i} 
=\alpha_{i} \textbf{B}_{i}~,
\end{eqnarray}     
where $\alpha_i$ is constant for a given $\textbf{B}_i$. The subfields
$\textbf{B}_1$ and $\textbf{B}_3$ are linear force-free having
$\alpha_1\neq\alpha_3$, whereas $\textbf{B}_2$ is a potential field
with $\alpha_2=0$. An optimal pair of $\alpha=\{\alpha_1,\alpha_3\}$
is iteratively found by minimizing the average deviation
between the observed transverse field ($\textbf{B}_t$) and the
computed ($\textbf{b}_t$) transverse field, quantified by
\begin{equation}
\label{En}
E_n=\left(\sum_{i=1}^{M} |\textbf{B}_{t,i}-\textbf{b}_{t,i}|\times |\textbf{B}_{t,i}|\right)/\left(\sum_{i=1}^{M}|\textbf{B}_{t,i}|^2\right)~,
\end{equation}  
on the photosphere. Here, $M=N^2$ represents
the total number of grid points on the transverse plane. The grid
points are weighted with respect to the strength of the observed
transverse field to minimize the contribution from weaker fields,
see \citep{HuDas08, Hu2010} for further details.

Since (\ref{tc}) involves the evaluation of the second-order
derivative, $(\nabla\times\nabla\times \textbf{B})_z=-(\nabla^2
\textbf{B})_z$ at $z=0$, evaluation of \textbf{B} requires magnetograms
at two different values of $z$. In order to work with the generally
available single-layer vector magnetograms, an algorithm was
introduced by \cite{Hu2010} that involves additional
iterations to successively fine-tune the potential subfield
$\textbf{B}_2$. The system is reduced to second order by  taking
initial guess $\textbf{B}_2=0$, which makes it easier to determine
the boundary condition for $\textbf{B}_1$ and $\textbf{B}_3$. If the 
calculated value of $E_n$ turns out unsatisfactory---i.e.,
overly large---then a potential field corrector to $\textbf{B}_2$
is calculated from the difference in the observed and computed
transverse fields and subsequently summed with the previous
$\textbf{B}_2$ to further reduce $E_n$. Notably, recent simulations
initiated with the non-FFF model have successfully explained the
circular ribbon-flares in AR NOAA 12192 \citep{avijeet2018} and AR
NOAA 11283 \citep{avijeet2020} as well as a blowout 
jet in AR NOAA 12615 \citep{Ss2019}, thus validating non-FFF's credibility.

The vector magnetogram is selected for 2019 March 08, at 03:00 UT
($\approx$ 7 minutes prior to the start of flare). The original
magnetogram cut out of dimensions 342$\times$195 pixels with pixel resolution 0.5 arcsec per pixel having an extent of $124~ \rm Mm\times 71$ Mm from
``hmi.sharp$\_$cea$\_$720s" series is considered, which ensures an
approximate magnetic flux balance at the bottom boundary. To
optimize the computational cost with the available resources, the
original field is re-scaled and non-FFF-extrapolated over a volume of
256$\times$128$\times$128 pixels while keeping the physical extent same
and preserving all magnetic structures throughout the region. The reduction, in effect, changes the conversion factor of 1 pixel to $\approx 0.484$ Mm along x and $\approx 0.554$ Mm along y and z directions of the employed Cartesian coordinate system.

\begin{figure}
\centering
\includegraphics[width=1\linewidth]{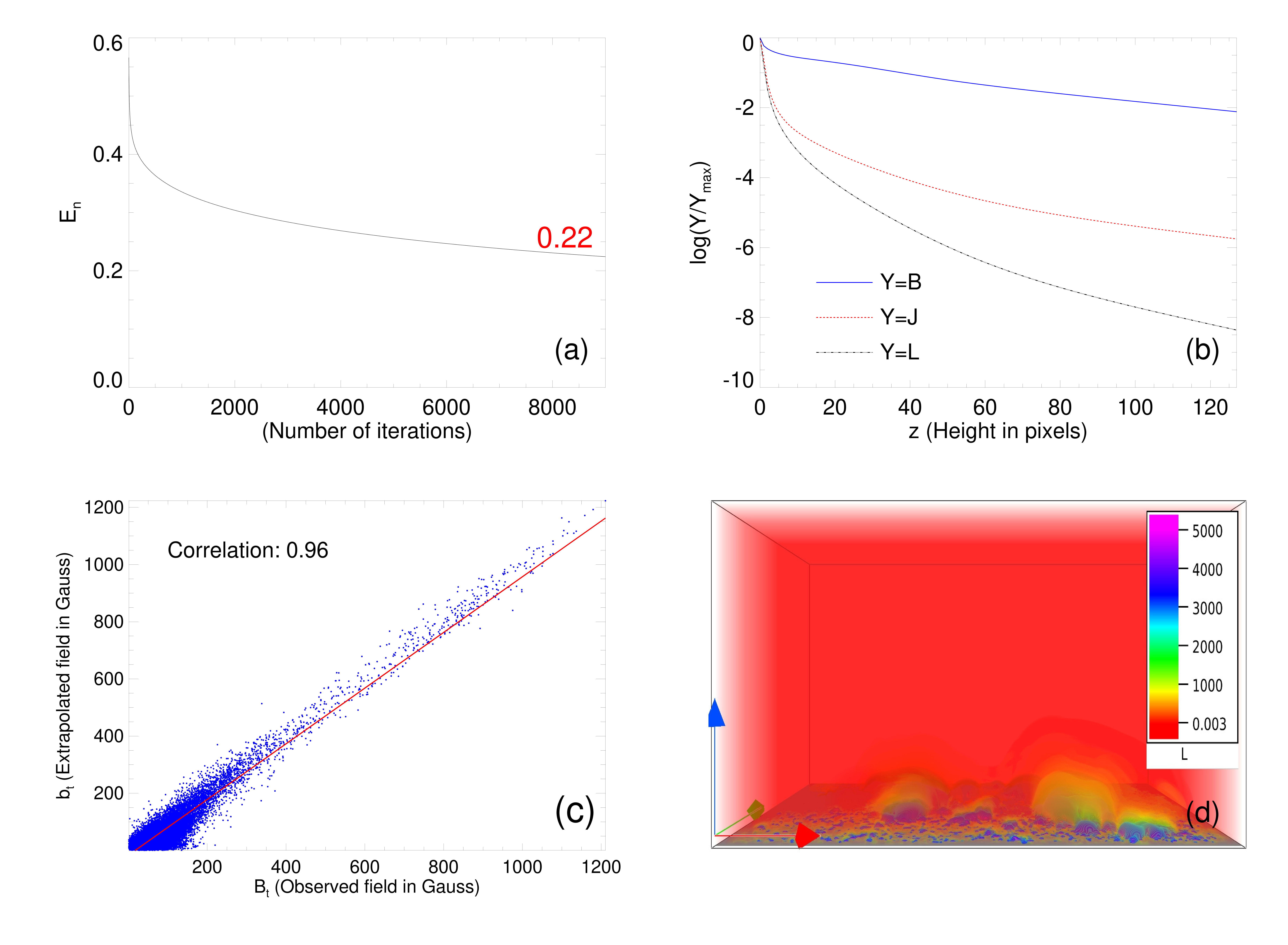}
\caption{Panel (a) shows the variation of the deviation $E_n$
with number of iterations in non-FFF extrapolation. Panel (b) shows
the logarithmic variation of horizontally averaged magnetic field
(Y=B), the current density (Y=J), and the Lorentz force 
(Y=L) with height $z$ in pixels. All the quantities plotted in panel
(b) are normalized with their respective maximum values. Panel (c)
shows the scatter plot of the correlation between the observed and
extrapolated magnetic field. The red line is the expected profile
for perfect correlation. Distribution of the magnitude of the
Lorentz-force for initial extrapolated field is shown in
panel (d) using direct volume rendering (DVR). The distribution
clearly shows that the Lorentz-force is maximum at bottom
boundary and decreasing with the height in computational volume.
The red, green and blue arrows on the bottom left corner represent
$x$, $y$ and $z-$ directions respectively here and hereafter. The
color bars on the right side of the panel represent the magnitude
of the strength of Lorentz-force .}
\label{lfcombnd}
\end{figure}

Panel (a) of Figure~\ref{lfcombnd} shows $E_n$ in the transverse 
field, defined in (\ref{En}), as a function of number of iterations.
It shows that $E_n$ tends to saturate at the value of $\approx$0.22.
Panel (b) of Figure \ref{lfcombnd} shows logarithmic decay of the
normalized horizontally averaged magnetic field, current density,
and Lorentz force with height. It is clear that the Lorentz force
is appreciable on the photosphere but decays off rapidly with height,
agreeing with the general perception that the corona is force-free
while the photosphere is not \citep{Liu2020, Yalim20}.  Panel (c)
shows that the Pearson-r correlation between the extrapolated and
observed transverse fields is $\approx$0.96, implying strong
correlation. The direct volume rendering of the Lorentz force in
panel (d) also reveals a sharp decay of the Lorentz force with
height, expanding on the result of panel~(b).

\begin{figure}
\centering
\includegraphics[width=0.8\textwidth]{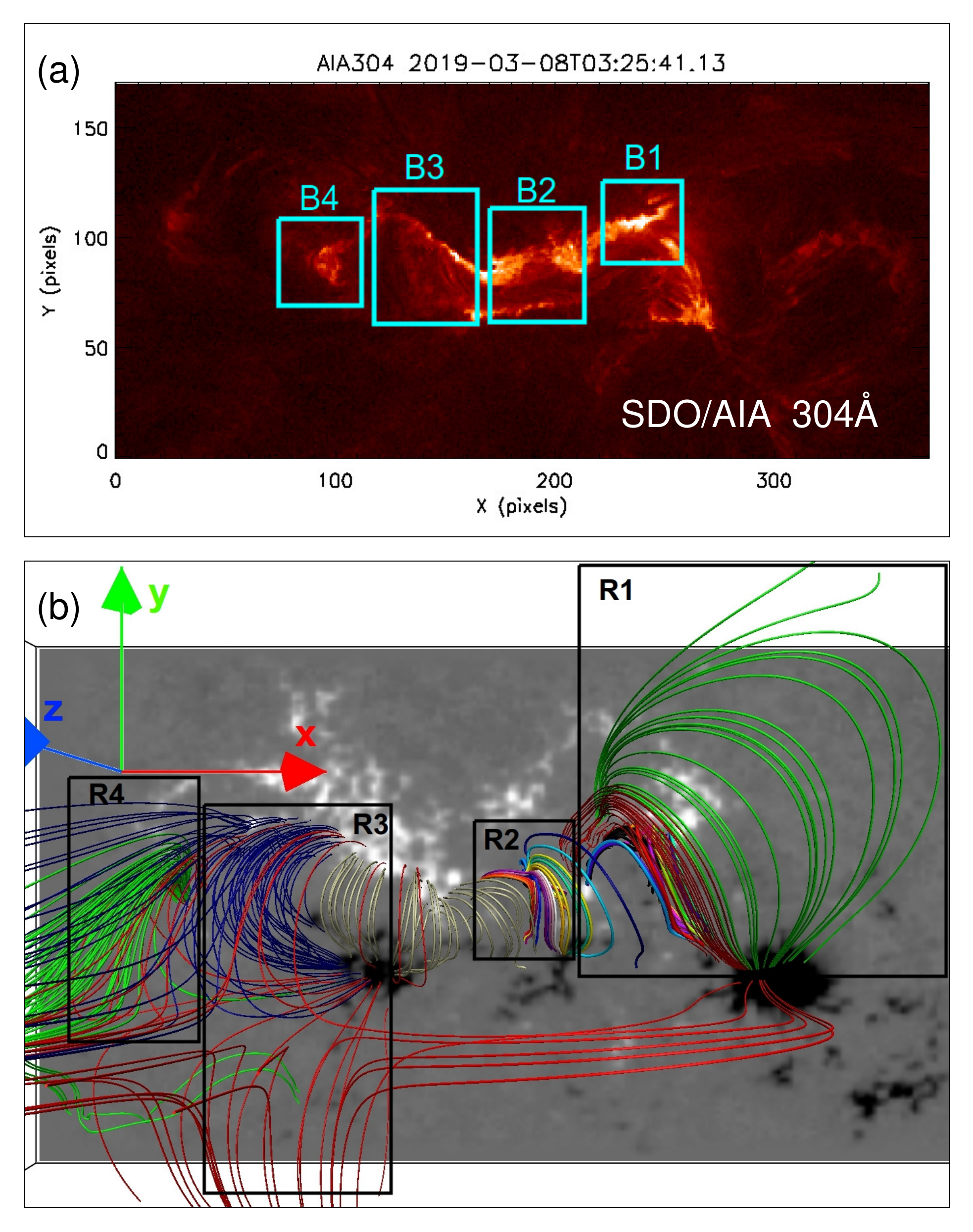}
\caption{Panel (a) shows the SDO/AIA 304{\AA} image where the flare
ribbon brightening has been divided into four parts B1, B2, B3, and
B4 (enclosed by boxes). Panel (b) shows an overall extrapolated
magnetic field lines morphology of AR NOAA 12734 with the
$B_{z}-$component of magnetogram at the bottom boundary. Foot points
of the magnetic structures contained in regions R1, R2, R3, and R4
correspond to the brightening B1, B2, B3, and B4 respectively.}
\label{regions}
\end{figure}

To facilitate description,  Figure \ref{regions}~(a) shows the
SDO/AIA 304 {\AA} image at 03:25 UT, where the flare ribbon brightening
has been divided into four segments marked as B1-B4. 
Figure \ref{regions}~(b) shows the initial global magnetic field
line morphology of AR NOAA 12734, partitioned into 
four regions R1-R4, corresponding to the flare ribbon brightening 
segments B1-B4. The bottom boundary of panel (b) comprises of
$B_z$ maps in grey scale where the lighter shade indicates positive
polarity regions and the darker shade marks the negative polarity
regions. The magnetic field lines topologies and structures
belonging to a specific region and contributing to the flare are
documented below. \bigskip

\noindent {\bf{Region R1:}} The top-down view of the global magnetic field
line morphology is shown in the panel (a) of Figure~\ref{region1}.
To help locate QSLs, the bottom boundary is overlaid 
with the $\log Q$ map of the squashing factor $Q$ \citep{Liu} in all 
panels of the figure. Distribution of high $Q$ values along with
$B_z$ on the bottom boundary helps in identifying differently
connected regions. The region with a large $Q$ is prone to the onset
of slipping magnetic reconnections \citep{Demoulin}.  Foot points
of magnetic field lines constituting QSL1 and QSL2 trace along the
high $Q$ values near the bottom boundary. QSL1, involving the
magnetic field lines Set I (green) and Set II (maroon), is shown
in panel (b). Particularly, magnetic field lines Set
I (green) extends higher in the corona forming the largest loops
in R1. Panel~(c) illustrates a closer view of QSL2
(multicolored) and the flux rope (black) beneath,
situated between the positive and negative polarities P1, P2 and
N1, respectively. In panel~(d), the flux rope (constituted by the
twisted black magnetic field lines) is depicted using the side view.
The twist value $T_w$ \citep{Liu} in the three vertical planes along the cross
section of the flux rope is also overlaid. Notably, the twist value
is 2 at the center of the rope and decreases outward (cf. vertical
plane in middle of the flux rope in panel (d)).  \bigskip

\noindent {\bf{Region R2:}} Figure~\ref{R2R3R4exp} (a) shows the
side view of a 3D null point geometry of magnetic
field lines and the bottom boundary $B_z$ overlaid
with log $Q$ ranging between 5 and 10. Panel~(b) depicts an enlarged
view of the 3D null location, marked black. The height of the null
is found to be $\approx$ 3~Mm from the photosphere. The null is
detected using the bespoke procedure \citep{DKRB, Ss2020} that
approximates the Dirac delta on the grid as 
\begin{equation}
\label{ndefine}
n(B_i) = \exp\big[-\sum_{i=x,y,z}{(B_{i} -B_{o})^2}/{d_{o}^2}\big]~, 
\end{equation}
where small constants $B_o$ and $d_o$ correspond to the isovalue
of $B_i$ and the Gaussian spread. The function $n(B_i)$ takes
significant values only if $B_i\approx 0~\forall i$, whereupon a
3D null is the point where the three isosurfaces having isovalues
$B_i=B_o$ intersect.\bigskip

\noindent {\bf{Region R3:}} Side view of the magnetic field line
morphology in region R3 is shown in Figure \ref{R2R3R4exp} (c),
where the yellow surface corresponds to $n=0.9$. Panel~(d) highlights 
a ``fish-bone-like'' structure, similar to the
schematic in Figure 5 of \citet{WangFB}. To show
that in the limiting case $n=0.9$ reduced to a null line, we plot
corresponding contours in the range $0.6\leq n \leq 0.9$ on three
pre-selected planes highlighted in panel (e). The size reduction
of the contours with increasing $n$ indicates the surface converging
to a line. Such null lines are also conceptualized as favorable
reconnection sites \citep{WangFB}.  \bigskip

\noindent  {\bf{Region 4}} Figure \ref{R2R3R4exp} (f) shows magnetic
field lines relevant to plasma rotation in B4. Notably, the null
line from the R3 intrudes into R4 and the extreme left plane in R3 (Figure \ref{R2R3R4exp} (e)) is also shared by the R4.

\section{HMHD and MHD simulations of AR NOAA 12734}
\label{simulation-results}
\subsection{Governing Equations and Numerical Model}
In the spirit of our earlier related works
\citep{avijeet2018, Ss2019, avijeet2020}, the plasma is idealized
to be incompressible and thermodynamically inactive as well as
explicitly nonresistive. While this relatively simple
idealization is naturally limited, it exposes the basic dynamics 
of magnetic reconnections unobscured by the effects due to 
compressibility and heat transfer. Albeit the latter are important 
for coronal loops \citep{2002ApJ...577..475R}, they do not directly 
affect the magnetic topology---in focus of this paper. Historically 
rooted in classical hydrodynamics, such idealizations have a proven
record in theoretical studies of geo/astrophysical phenomena
\citep{Rossby38, 1991ApJ...383..420D, RBCLOW, 2021ApJ...906..102B}.
Inasmuch as their cognitive value depends on an a posteriori validation
against the observations, the present study offers yet another  
opportunity to do so.

The Hall forcing has been incorporated \citep{2021ApJ...906..102B} 
in the computational model EULAG-MHD \citep{PiotrJCP} to solve the 
dimensionless HMHD equations,
\begin{eqnarray}
\label{momtransf}
\frac{\partial{\bf v}}{\partial t} +({\bf v}\cdot \nabla){\bf v}&=&
 -\nabla p + (\nabla\times{\bf B})\times{\bf B} + 
\frac{1}{R_F^A}\nabla^2 {\bf v}~,\\
\label{induc}
\frac{\partial{\bf B}}{\partial t}&=& \nabla\times(\textbf{v}\times{\bf B})
-d_H\nabla\times((\nabla\times{\bf B})\times{\bf B})~,\\
\label{incompv}
\nabla\cdot {\bf v}&=& 0~, \\
\label{incompb}
\nabla\cdot {\bf B}&=& 0~,
\end{eqnarray}
where $R_F^A=(v_A L/\nu)$, $\nu$ being the kinematic viscosity---is an effective fluid Reynolds number, 
having the plasma speed replaced by the Alfv\'en  speed  $v_A$.
Hereafter $R_F^A$ is denoted as fluid Reynolds number for convenience. The transformation of the dimensional quantities (expressed in cgs-units) 
into the corresponding non-dimensional quantities,
\begin{equation}
 \label{norm}
       {\bf{B}}\longrightarrow \frac{{\bf{B}}}{B_0},
  \quad{\bf{x}}\longrightarrow \frac{\bf{x}}{L_0},
  \quad{\bf{v}}\longrightarrow \frac{\bf{v}}{v_A},
  \quad t \longrightarrow \frac{t}{\tau_A},
  \quad p \longrightarrow \frac{p}{\rho_0 {v_{A}}^2}~, 
 \end{equation}
assumes arbitrary $B_0$ and $L_0$ while the Alfv\'en speed $v_A \equiv
B_0/\sqrt{4\pi\rho_0}$. Here $\rho_0$ is a constant mass density,
and $d_H$ is the Hall parameter. In the limit of $d_H=0$,
(\ref{momtransf})-(\ref{incompb}) reduce to the MHD equations
\citep{avijeet2018}.

The governing equations (\ref{momtransf})-(\ref{incompb})
are numerically integrated using EULAG-MHD---a magnetohydrodynamic
extension \citep{PiotrJCP} of the established Eulerian/Lagrangian
comprehensive fluid solver EULAG \citep{Prusa08} predominantly used 
in atmospheric research. The EULAG solvers are based on the
spatio-temporally second-order-accurate nonoscillatory forward-in-time
advection scheme MPDATA (for {\it multidimensional positive definite
advection transport algorithm}) \citep{Piotrsingle}.  Importantly, 
unique to MPDATA is its widely
documented dissipative property mimicking the action of explicit
subgrid-scale turbulence models wherever the concerned advective
field is under-resolved; the property known as implicit
large-eddy simulations (ILES) \citep{Grinstein07}. In effect, 
magnetic reconnections resulting in our simulations dissipate the 
under-resolved magnetic field along with other advective
field variables and restore the flux freezing. These reconnections
being intermittent and local, successfully mimic physical reconnections.

\subsection{Numerical Setup}
The simulations are carried out by mapping the physical domain of $256\times128\times128$ pixels on the computational domain of $x\in\{-1, 1\}$, $y\in\{-0.5,0.5\}$, $z\in\{-0.5,0.5\}$ in a Cartesian coordinate system. The dimensionless spatial step sizes are $\Delta x=\Delta y=\Delta z \approx 0.0078$. The dimensionless time step is  $\Delta t=5\times 10^{-4}$, set to resolve whistler speed---the fastest
speed in incompressible HMHD. The rationale is briefly presented in the Appendix \ref{appnd}.
The corresponding initial state is motionless ($\textbf{v}=0$) and the initial
magnetic field is provided from the non-FFF extrapolation. The non-zero
Lorentz force associated with the extrapolated field pushes the
magnetofluid to initiate the dynamics. Since the maximal variation
of magnetic flux through the photosphere is only 2.28$\%$ of its
initial value during the flare (not shown), the $\text{B}_z$ at the
bottom boundary (at $z=0$) is kept fixed throughout the simulation
 while all other boundaries are
kept open. For velocity, all boundaries are set open. The mass density is set to $\rho_0=1$.  

 The fluid Reynolds number is set to $500$, which is roughly two orders of magnitude smaller than its coronal value $\approx 25000$ (calculated using kinematic viscosity $\nu=4\times 10^9 ~\rm m^2s^{-1}$ \citep{Aschwanden} in solar corona).
 Without any loss in generality, the reduction in $R_F^A$ can be envisaged
to cause a reduction in computed Alfv\'en speed, $v_A|_\text{computed} \approx 0.02\times v_A|_\text{corona}$ where the $L$ for the computational and coronal length scales are set to 71 Mm and 100 Mm respectively. This diminished Alfv\'en speed reduces the requirement of computational resources and also relates it with the observation time. The results presented herein pertain to a run for 1200$\Delta t$ which along with the normalizing $\tau_A\approx 3.55\times 10^3$ s roughly corresponds to an observation time of $\approx$ 35 minutes. For the ease of reference in comparison with observations, we present the time in units of 0.005$\tau_a$ (which is 17.75 s) in the discussions of the figures in subsequent sections.

Although the  coronal plasma idealized to have reduced Reynolds number is inconsequential here, in a comparison of MHD and HMHD evolution, we believe the above rationale merits further contemplation. Undeniably such a  coronal plasma is not a reality. Nevertheless, the reduced $R_F^A$ does not affect the reconnection or its 
consequence, but slows down the dynamics between two such events and importantly---reduces the computational cost, making data-based simulations realizable even with reasonable computing resources. 
A recent work by \citet{JiangNat} used homologous approach toward simulating a realistic and self-consistent flaring region.

In the present simulations, all
parameters are identical for the MHD and the HMHD
except for the $d_H$, respectively set to 0 and 0.004.
The value 0.004 is motivated by recognizing ILES dissipation models intermittent magnetic reconnections at the ${\mathcal O}(\parallel\Delta{\bf x}\parallel)$ length scales,
consistent with the thesis put forward in Introduction, we specify
an appreciable Hall coefficient as $d_H = 0.5 \Delta z/L \approx
0.004$, where $L=1\equiv$ smallest extent of the computational volume, 
having $\Delta y= \Delta z \approx 0.0078$ as the dissipation scales because of the ILES
property of the model. Correspondingly, the value is also at the lower bound of the pixel or scale order
approximation and, in particular, an order of magnitude smaller
that its coronal value valid at the actual dissipation scale.  An
important practical benefit of this selection is the optimization
of the computational cost while keeping magnetic field line dynamics
tractable. Importantly, with dissipation and Hall scales being tied, an increased current density at the dissipation scale introduces additional slippage of field lines in HMHD over MHD (due to the Hall term) and, may be responsible for more effective and faster reconnections found in the Hall simulation reported below.

 \subsection{Comparison of the HMHD and MHD simulations}
The simulated HMHD and MHD dynamics leading to the flare show
unambiguous differences. This section documents these differences
by comparing methodically simulated evolution of the magnetic
structures and topologies in the AR NOAA 12734---namely,
the flux rope, QSLs, and null points---identified in the extrapolated
initial data in the regions R1-R4.

\subsubsection{Region R1} 
The dynamics of region R1 are by far the most complex among the 
four selected regions. To facilitate future reference as well as to outline the 
organization of the discussion that follows, Table~\ref{tab:r1} provides a brief 
summary of our findings---in a spirit of theses to be proven by the simulation results.
\begin{table}
\caption{Salient features of magnetic field lines dynamics in R1}
\label{tab:r1}
\begin{tabular}{ |p{3cm}|p{5.5cm}|p{5.5cm}|  }
\hline
Magnetic field lines structure&   HMHD  &   MHD \\ [4ex]
\hline
QSL1 & Fast reconnection followed by a significant rise of loops,
eventually reconnecting higher in the corona. &Slow reconnection
followed by a limited rise of loops. \\ [6ex]
\hline
QSL2 & Fast reconnection causing the magnetic field lines to entirely
disconnect from the polarity P2.  & Due to slow reconnection magnetic
field lines remain connected to P2. \\ [6ex]
\hline
Flux rope &Fast slipping reconnection of the flux-rope foot points,
followed by the expansion and rise of the rope envelope.  & Slow
slipping reconnection and rise of the flux-rope envelope; the
envelope does not reach the QSL1. \\ [6ex]
\hline
\end{tabular}
\end{table}
\bigskip

\begin{figure}
\centering
\includegraphics[width=1\textwidth]{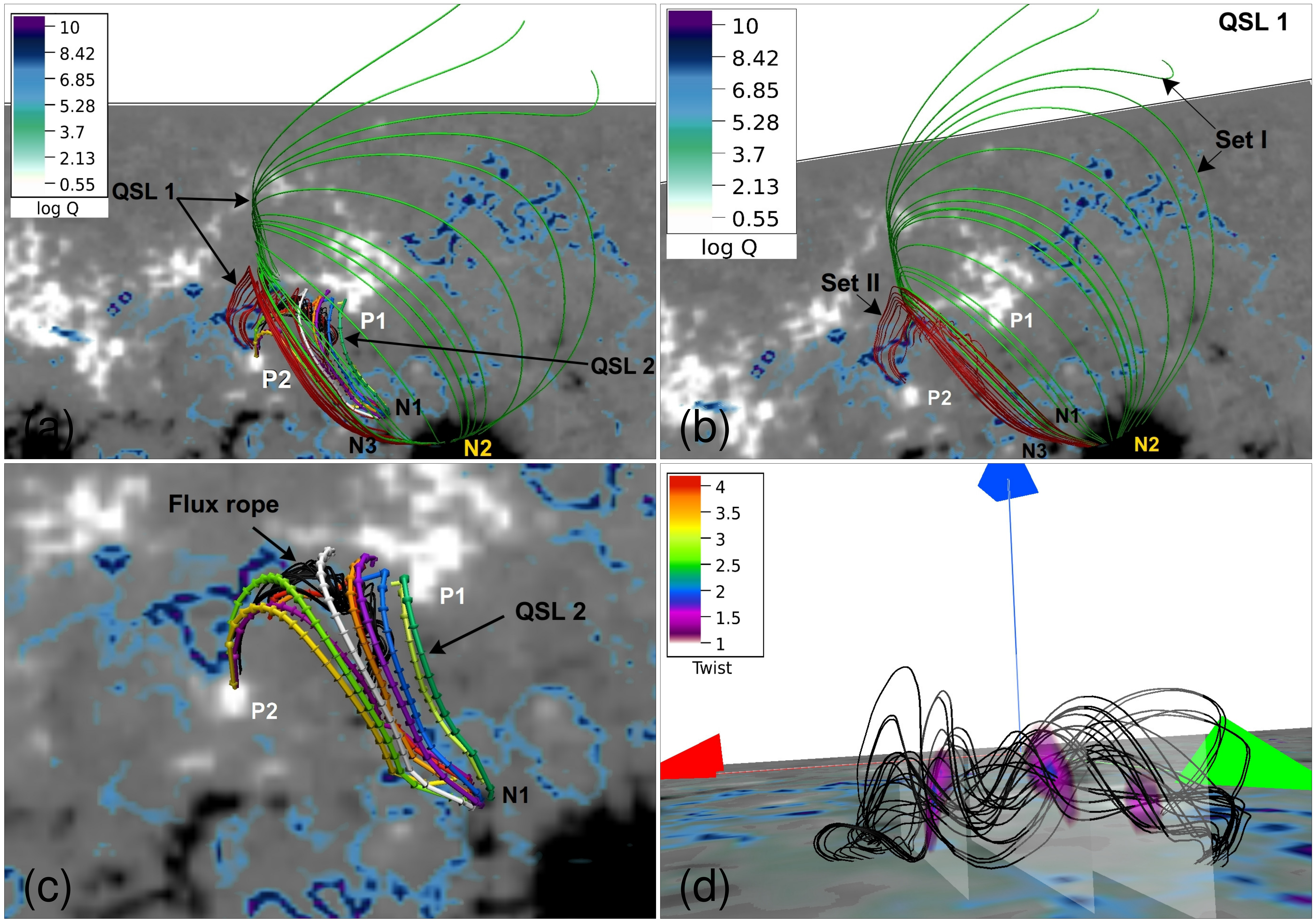}
\caption{Panel (a) shows magnetic field lines morphology of region R1 between positive and negative polarities P1, P2, N1, N2, and N3 respectively. Panel (b) highlights the structure of QSL 1 comprised of magnetic field lines Set I (green) and Set II (maroon). Panel (c) shows the zoomed top view of the flux rope structure (black) and an overlying QSL 2 (multi color arrowed magnetic field lines), between the positive and negative polarities P1, P2, and N1 respectively. Panel (d ) shows the side view of the flux rope where three vertical planes along the cross section of the flux rope show the twist value $T_w$ at different locations along the flux rope. In all the panels the log $Q$ between 5 and 10, is overlaid on $B_{z}-$component of magnetogram at the bottom boundary.}
\label{region1}
\end{figure}

\begin{figure}
\centering
\includegraphics[width=0.9\textwidth]{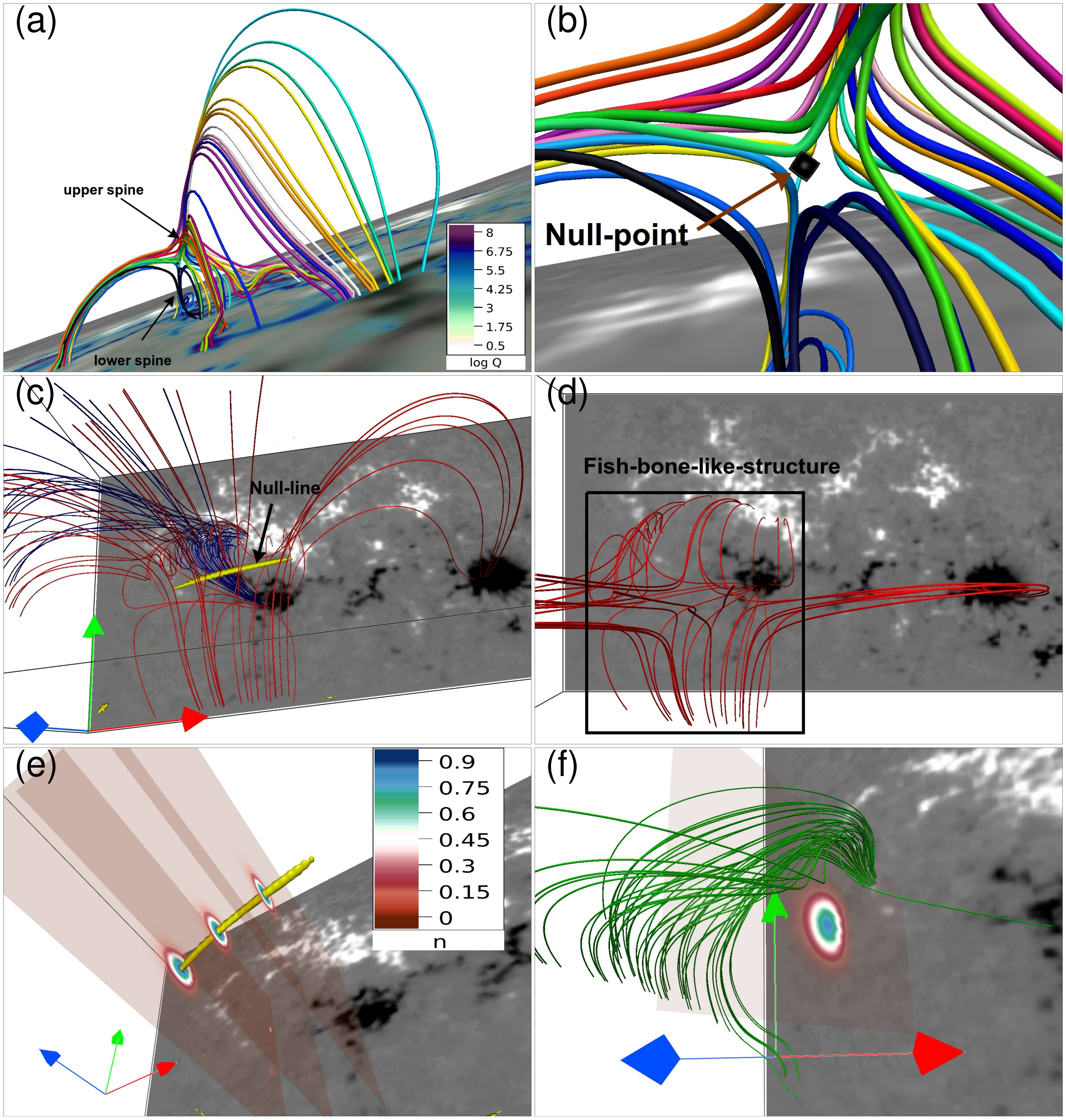}
\caption{Panel (a) shows a 3D null spine-fan configuration in region R2 with the $B_z$ as the bottom boundary overlaid with log $Q$ between 5 and 10. Panel (b) is zoomed view of (a), highlighting the 3D null-point (in black)---an iso surface of $n=0.6$ indicated by an arrow. Panel (c) shows the side view of magnetic field lines structure in region R3 along with the yellow surface representing the null-line corresponding to $n=0.9$. Panel (d) shows the top-down view of red magnetic field lines of (c) forming fish-bone-like structure. In panel (e) we show the value of $n$ on the three different  vertical planes passing through the cross-sections of the null-line surface. Notably, the planes show circles in the cross-section at different locations which indicates that the yellow surface is a null line. Panel (f) depicts magnetic field lines morphology in region R4 along with the value of $n$ on a vertical plane where the green circular contour corresponds to $n=0.6$ suggesting the right part of magnetic field lines morphology may be a part of the null-line geometry (shown in panel (c)).} 
\label{R2R3R4exp}
\end{figure}

\begin{figure}
\centering
\includegraphics[width=1\linewidth]{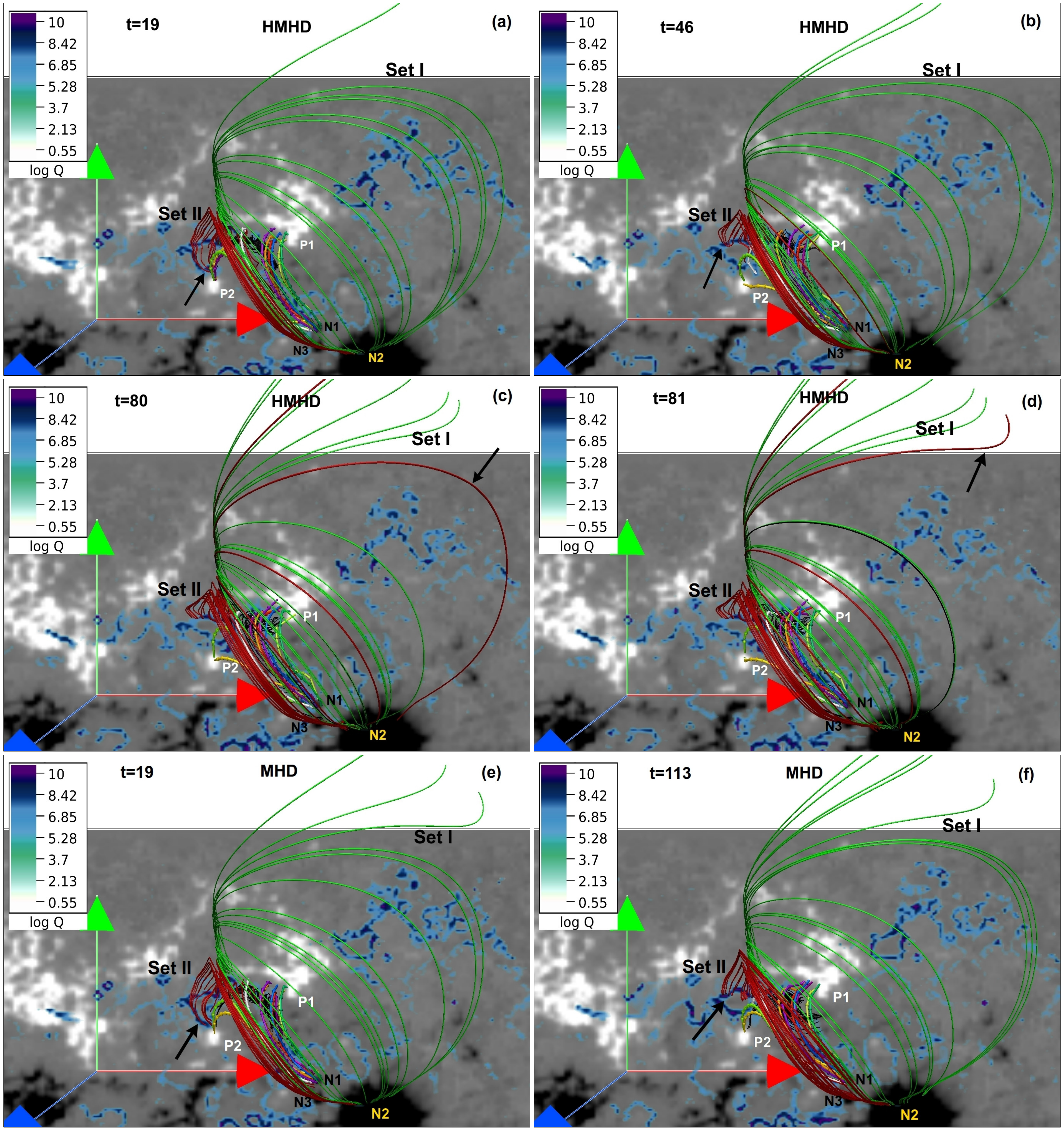}
\caption{Snapshots of the global dynamics of magnetic field lines in  region R1 during the HMHD and MHD simulations are shown in panels (a)-(d) and panels (e)-(f) respectively. Panels (a) and (b) show the departure of foot points of magnetic field lines Set II (maroon) away from polarity P2 between t=19 and 46 on the bottom boundary (marked by black arrow). Panel (c) depicts the rising magnetic field lines Set II (maroon) higher up in the solar corona at t=80 (marked by black arrow) and (d) shows subsequent  connectivity change of rising magnetic field lines at t= 81, due to reconnection with the Set I (green) magnetic field lines. Panels (e) and (f) depict the departure of foot points of magnetic field lines Set II (maroon) away from P2 between t= 19 to 113 which is similar to the HMHD but delayed in time---indicating the slower dynamics in the MHD. Notably, significant rise of magnetic field lines Set II and consequent reconnection of it with magnetic field lines Set I higher up in the solar corona is absent in the MHD simulation. An animation of this figure is available. The video shows the  evolution of magnetic field lines in region R1 from t=0 to 109 for the HMHD and from t=0 to 120 for the MHD simulations respectively. The realtime duration of the video is 12 s.}
\label{fullR1}
\end{figure}

\begin{figure}
\centering
\includegraphics[width=1\linewidth]{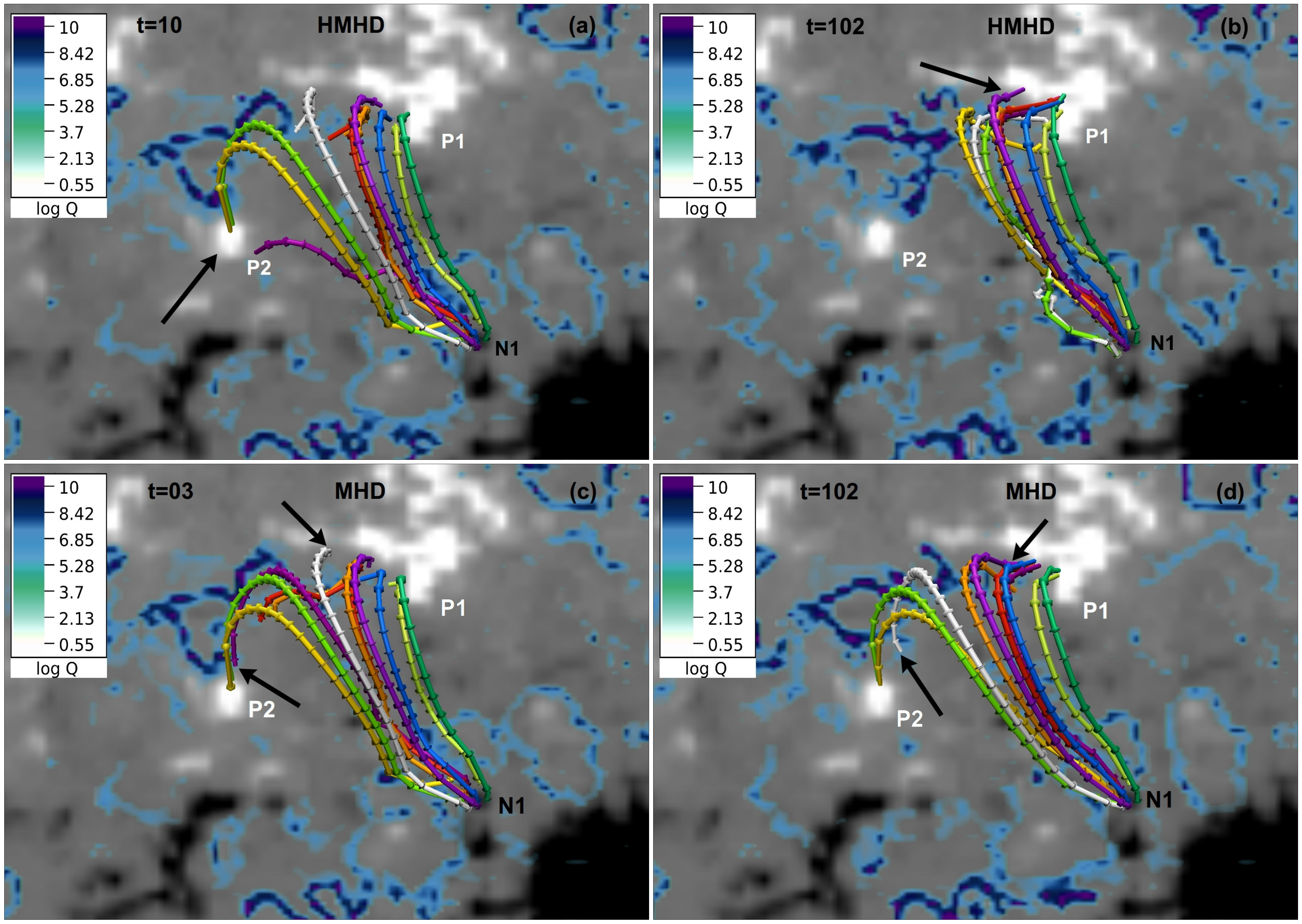}
\caption{Snapshots of the HMHD and MHD evolution of QSL2 (Figure \ref{region1} (c)) are shown in panels (a)-(b) and panels (c)-(d) respectively. Panels (a)-(b) show magnetic field linesanchored in the positive polarity P2 at t=10 have moved to the polarity P1 by t=102 and changed their connectivity (marked by black arrow) due to reconnection along QSL during the HMHD. Panels (c)-(d) show the connectivity changes of the violet and white color magnetic field lines during the MHD evolution. The white field line was initially connecting the polarities P1 and N1 whereas the violet field line was connecting P2 and N1. As a result of reconnection along QSL the white field line changes its connectivity from P1 to P2 and violet field line changes the connectivity from P2 to P1 (marked by black arrows). Notably, unlike the HMHD simulation not all magnetic field lines move to P1 from P2 due to reconnection along QSL during the MHD which indicates the slower dynamics. An animation of this figure is available. The video shows the evolution of QSL2 (shown in Figure \ref{region1}) in region R1 from t=0 to 120 for the HMHD and MHD simulations respectively. The realtime duration of the video is 12 s.}
\label{R1QSL}
\end{figure}

\begin{figure}
\centering
\includegraphics[width=1\textwidth]{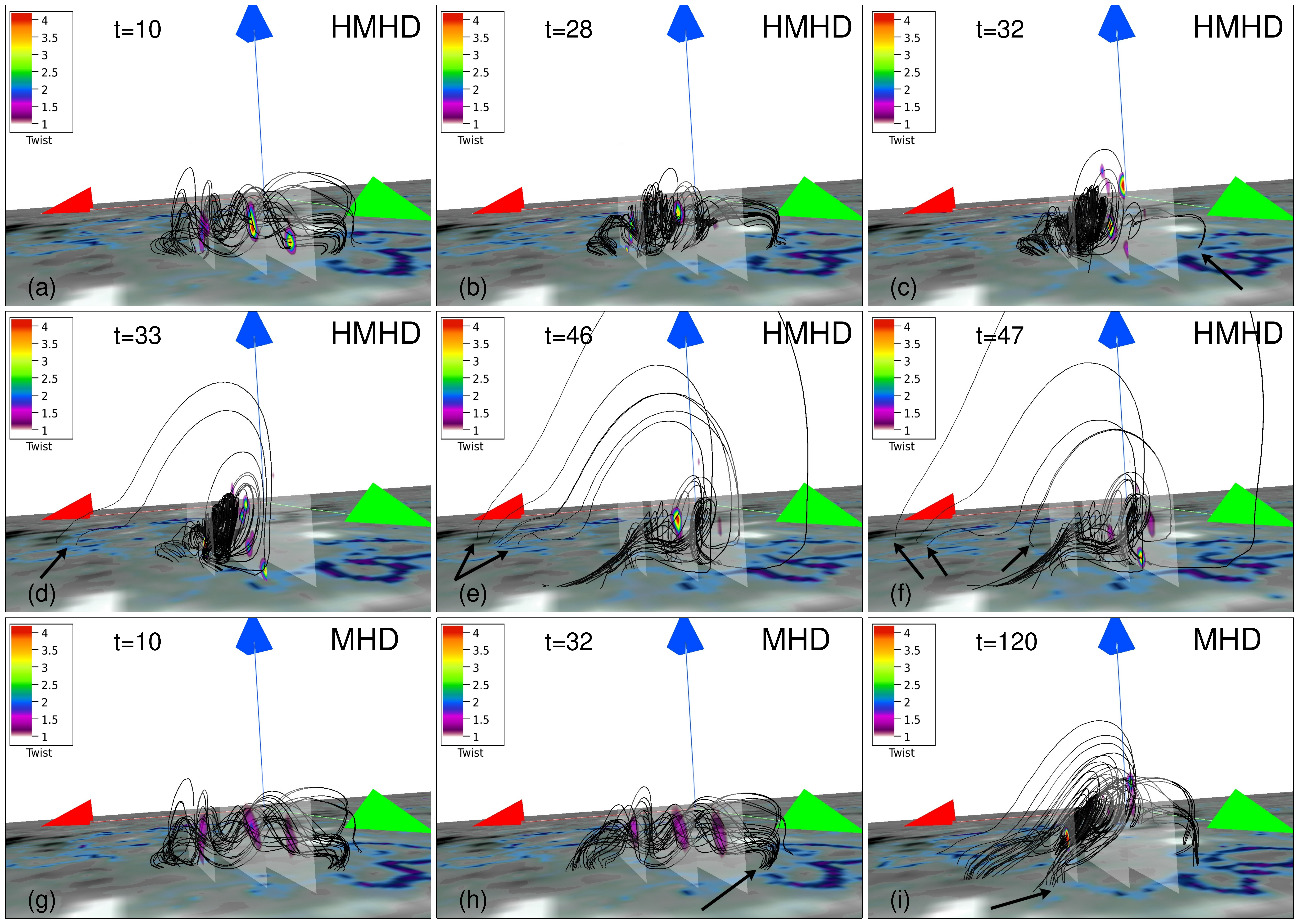}
\caption{Time sequence showing the HMHD (panels (a)-(f)) and MHD (panels (g)-(i)) evolution of the flux rope (shown in Figure \ref{region1}(c)) along with the twist $T_w$. Panel (a) shows the twist on the middle and the right plane on flux rope is higher than initial values (c.f. Figure \ref{region1}(d)) and reduced with time in panel (b). Panels (c)-(d) depict the connectivity change of the foot point of rope from right to left (indicated by black arrow) due to reconnection along QSL. Panels (e)-(f) show the connectivity change of magnetic field lines on left hand side (indicated by black arrow). Panels (g)-(i) depict the dynamic rise of the flux rope between t=10 and t=120 during the MHD simulation. Notably, the foot points of the rope on the right side (marked by black arrow) at t=32 in (h) have moved towards left by t=120 in (i) as a result of reconnection along QSL. An animation of this figure is available. The video shows the evolution of magnetic flux rope in region R1 from t=0 to 92 for the HMHD and from t=0 to 120 for the MHD simulations respectively. The realtime duration of the video is 12 s.}
\label{ropeHMHD-MHD}
\end{figure}

The global dynamics of magnetic field lines in region R1 is
illustrated in Figure~\ref{fullR1}; consult
Figure~\ref{region1} for the initial condition and terminology. The
snapshots from the HMHD and MHD simulations are shown in panels
(a)-(d) and (e)-(f), respectively. In panels (a) and (b), corresponding
to $t=19$ and $t=46$, the foot points of magnetic field lines Set
II (near P2, marked maroon) exhibit slipping reconnection along
high values of the squashing factor $Q$ indicated by black arrows.
Subsequently, between $t=80$ and 81 in panels (c) and (d), the
magnetic field lines Set II rise in the corona and reconnect with
magnetic field lines Set I to change connectivity. The MHD counterpart
of the slipping reconnection in panels (e) and (f), corresponds to
magnetic field lines Set II between t=19 and t=113. It lags behind
the HMHD displays, thus implying slower dynamics. Furthermore, the
magnetic field lines Set II, unlike for the HMHD, do not reach up
to the magnetic field lines Set I constituting QSL1 and hence do
not reconnect.  A more informative visualization of the highlighted
dynamics is supplemented in an online animation. The decay index is calculated for each time instant for both the simulations and is found to be less than 1.5 above the flux rope, indicating an absence of the torus instability \citep{Torok}.
For more detail,
Figures~\ref{R1QSL} and \ref{ropeHMHD-MHD} illustrate evolution of
QSL2 and flux rope separately.

Figure~\ref{R1QSL} panels (a)-(b) and (c)-(d) show,
respectively, the instants from the HMHD and MHD simulations of
QSL2 between P1, P2 and N1. The HMHD instants show
magnetic field lines that were anchored between P2
and N1 at $t=10$ have moved to P1 around t=102, marked by black
arrows in both panels.  The magnetic field lines anchored at P2
moved to P1 along the high $Q$ values---signifying the slipping
reconnection. The MHD instants in panels (c)-(d)
show the connectivity changes of the violet and white colored
magnetic field lines. The white field line was initially connecting
P1 and N1, whereas the violet field line was connecting P2 and N1.
As a result of reconnection along QSL, the white field line changed
its connectivity from P1 to P2 and violet field line changes the
connectivity from P2 to P1 (marked by black arrows). Notably, in
contrast to the HMHD evolution, all magnetic field lines initially
anchored in P2 do not change their connectivity from P2 to P1 during
the MHD evolution, indicating the slower dynamics.

The flux rope has been introduced in panels (c) and
(d) of Figure~\ref{region1}, respectively, below the QSL2 and in
enlargement. Its HMHD and MHD evolutions along with the twists on
three different vertical cross sections are shown in panels (a)-(f)
and (g)-(i) of Figure~\ref{ropeHMHD-MHD}, respectively.  Magnetic
field lines constituting the rope, rise substantially higher during
the HMHD evolution as a result of slipping reconnection along the high $Q$
in panels (c)-(f). In panel (c) at $t=32$, the foot points of the
rope that are anchored on right side (marked by black arrow) change
their connectivity from one high $Q$ regime to another in panel (d)
at t=33; i.e., the foot points on the right have moved to the left
side (marked by black arrow).  Afterwards, the magnetic field lines rise because of the
continuous slipping reconnection, as evidenced in panels (e) to (f)
and the supplemented animation.  Comparing panels (a) with (g) at 
$t=10$ and (c) with (h) at t=32, we note that the twist
value $T_w$ is higher in the HMHD simulation. Panels 
(h)-(i) highlight the displaced foot points of flux rope due to slipping reconnection
at t=32 and t=120 (cf. black arrow). The rope is preserved throughout the 
HMHD and MHD simulations.

\begin{figure}
\centering
\includegraphics[width=1\textwidth]{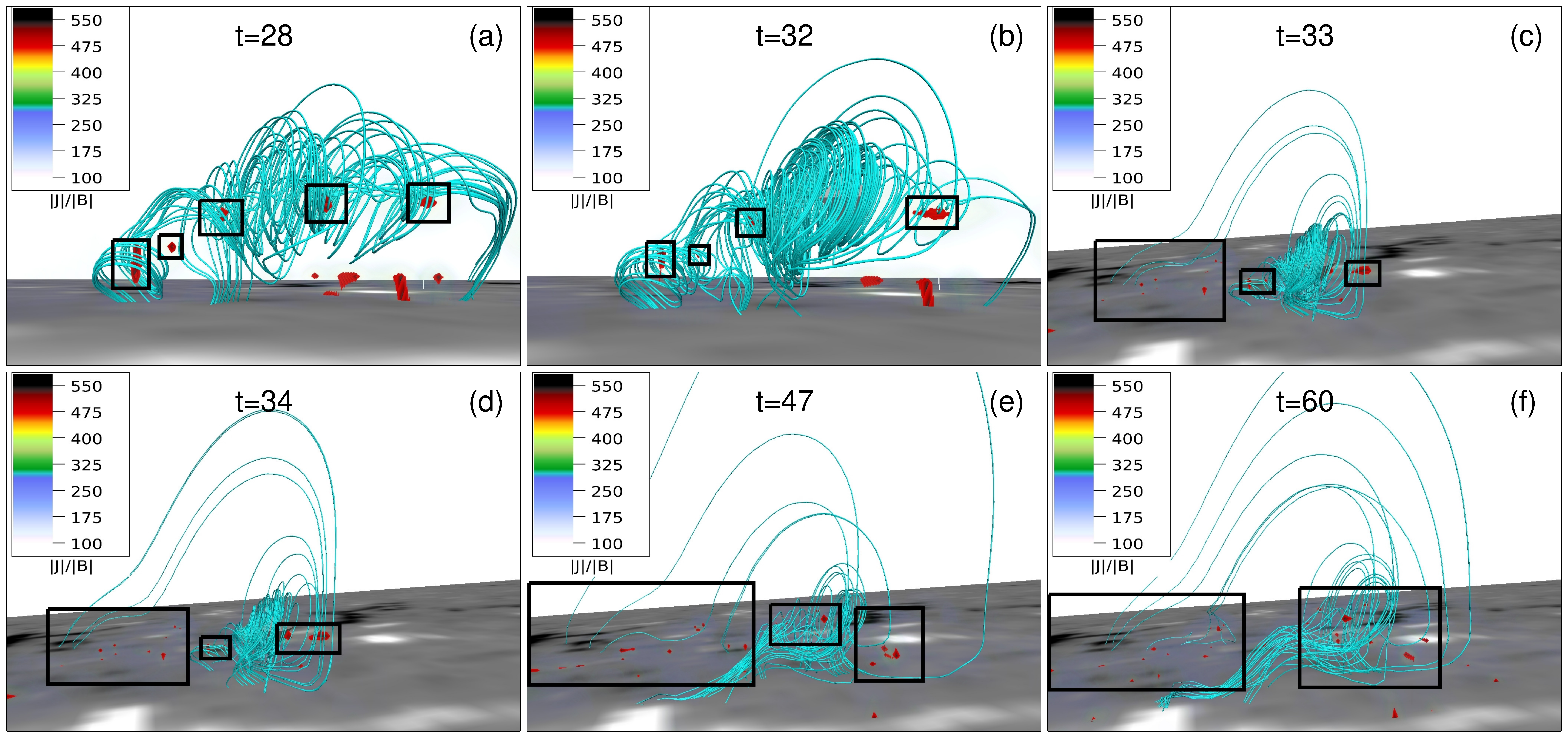}
\caption{Temporal variation of the direct volume rendering of  $(|\textbf{J}|/|\textbf{B}|)$ along with the flux rope is shown during the HMHD simulation. Noticeably, the high magnetic field gradient regions with $(|\textbf{J}|/|\textbf{B}|)\ge 475$ develop within (panels (a) to (c)) and on the left side of the flux rope (panels (d) to (f)). The values $(|\textbf{J}|/|\textbf{B}|)\ge 475$ in each panel are enclosed within the black rectangular boxes.}
\label{ropecs}
\end{figure}

\begin{figure}
\centering
\includegraphics[width=1\textwidth]{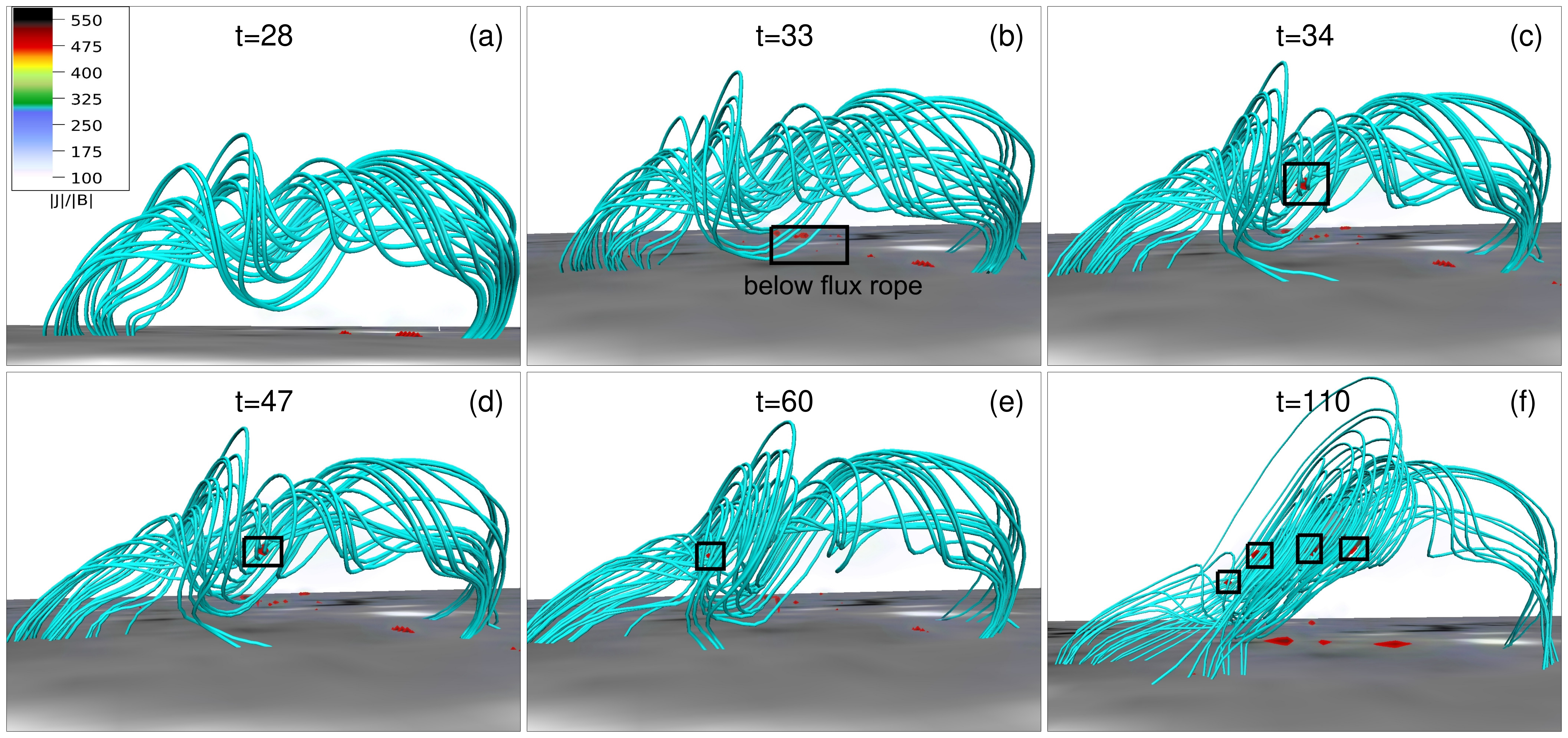}
\caption{Temporal variation of the direct volume rendering of $(|\textbf{J}|/|\textbf{B}|)$ along with the flux rope is shown during the MHD simulation. Panels (a) and (b) shows the absence of high values of $(|\textbf{J}|/|\textbf{B}|)$ within the rope at $t=28$ and $t=32$ but in later panels (c) to (f) $(|\textbf{J}|/|\textbf{B}|)\ge475$ appears (enclosed by the black rectangular boxes). Notably, as compared to the HMHD case (Figure \ref{ropecs}), the development of $(|\textbf{J}|/|\textbf{B}|)$ is not significant in the region R1 during the MHD.}
\label{ropecsmhd}
\end{figure}

The rise and expansion of the flux-rope envelope
owing to slipping reconnection is remarkable in the
HMHD simulation.  \citet{dudik} have already shown such a flux-rope
reconnection along QSL in a J-shaped current region,
with slipping reconnection causing the flux rope to form a sigmoid
(S-shaped hot channel observed in EUV images of SDO/AIA) followed
by its rise and expansion. Further insight is gained by overlaying 
the flux rope evolution shown in Figure \ref{ropeHMHD-MHD} with direct volume rendering of
 $|{\bf J}|/|{\bf B}|$ (Figures \ref{ropecs} and \ref{ropecsmhd}) as a measure of magnetic field gradient  for the HMHD and MHD simulations.  
In the HMHD case, appearance of large values of $|{\bf J}|/|{\bf B}|>475$ inside the rope 
(panels (a) to (c)) and foot points on left of the rope (panels (d) to (e)) are apparent. 
The development of the large $|{\bf J}|/|{\bf B}|$ is indicative of reconnection 
within the rope. Contrarily, MHD simulation lacks such high values of $|{\bf J}|/|{\bf B}|$
in the same time span (panels (a)-(b)) and the field lines show no slippage---agreeing with the proposal that large currents magnify the Hall term, resulting into more effective slippage of field lines.

\subsubsection{Region R2} 
\begin{figure}
\centering
\includegraphics[width=1\textwidth]{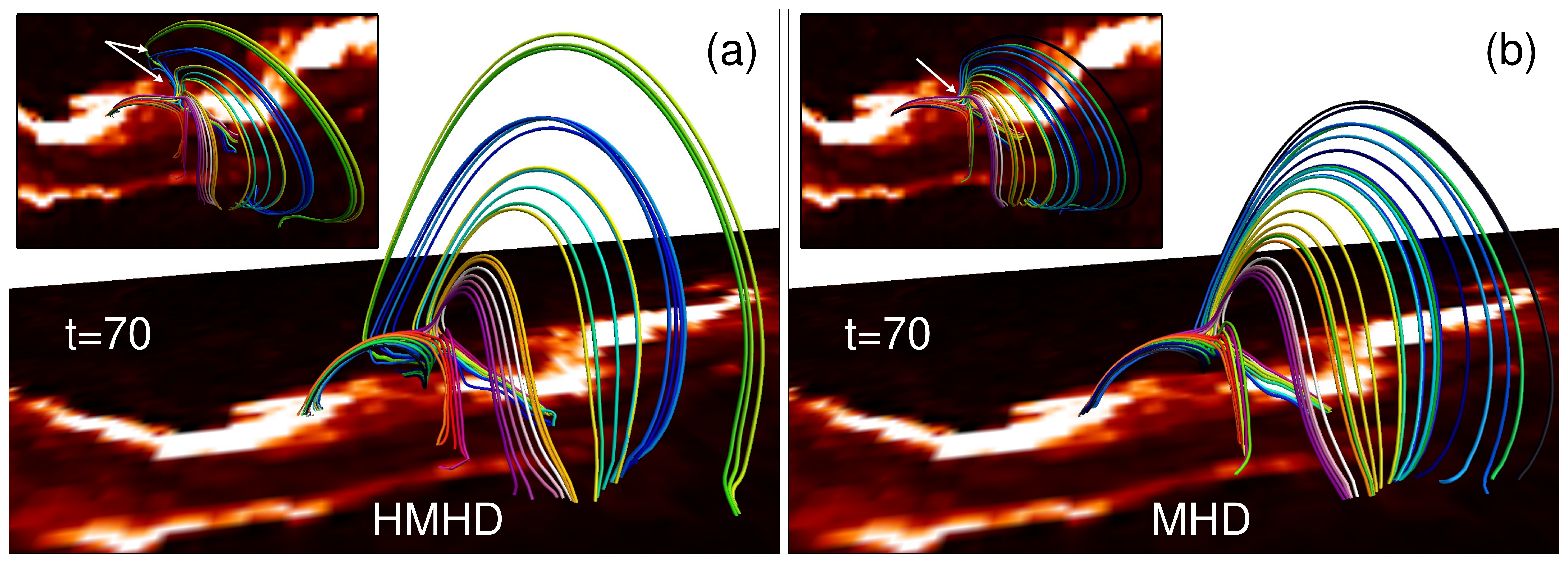}
\caption{Panels (a) and (b) show the comparison of magnetic field lines topology in region R2 at t=70 with the flare ribbons observed in the SDO/AIA 304 {\AA} channel (side views) during the HMHD and MHD simulations respectively. The inset images on the top left corner in each panel show the top view of the same magnetic field lines topology. Notably, the spine is anchored in the HMHD while it is not connected to the bottom boundary in the MHD at t=70 (marked by white arrow in inset images). An animation of this figure is available. The video shows the evolution of magnetic field lines in region R2 with the AIA 304 {\AA} image at the bottom boundary from t=0 to 117 for the HMHD and from t=0 to 120 for the MHD simulations respectively. The realtime duration of the video is 12 s.}
\label{R2comp}
\end{figure}

\begin{figure}
\centering
\includegraphics[width=1\textwidth]{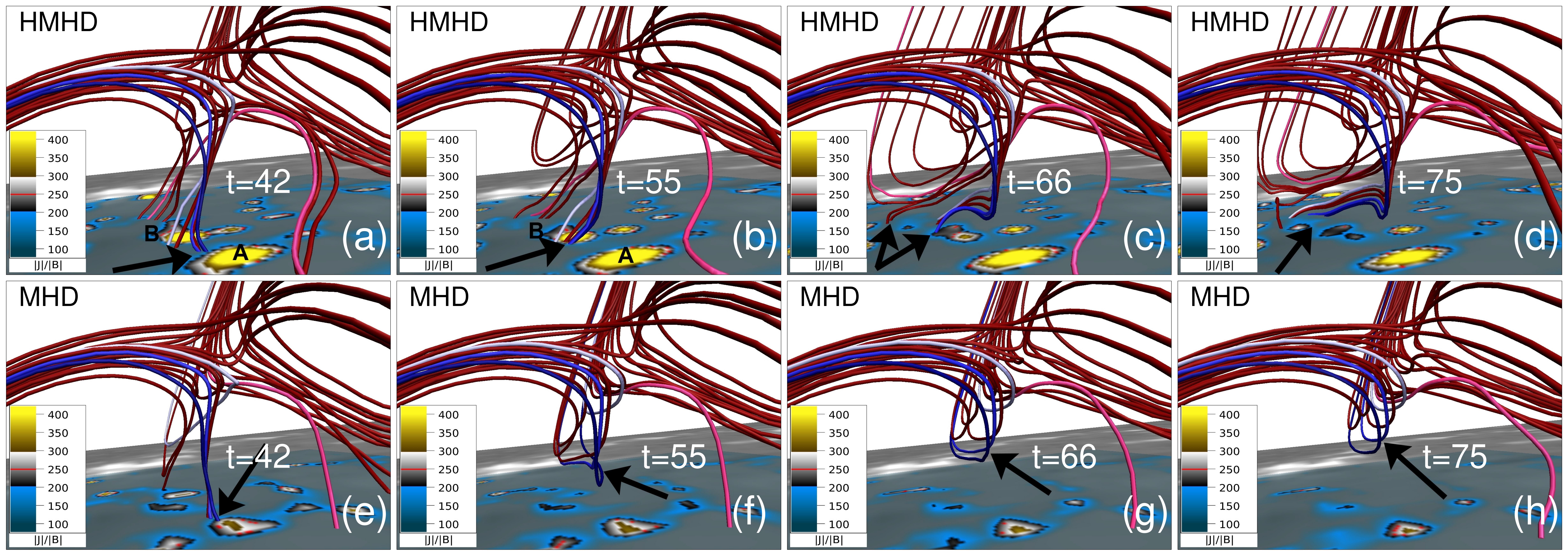}
\caption{Panels (a) to (d) depict the slipping motion of the lower spine field lines (also shown in the Figure \ref{R2R3R4exp}) overlaid with the $|\textbf{J}|/|\textbf{B}|$ on the bottom boundary during the HMHD evolution. The motion is marked by the black arrows in all the panels indicating the successive change in the location of field lines on the bottom boundary. A and B (in panels (a) and (b)) are the two regions with $|\textbf{J}|/|\textbf{B}|>350$ on the bottom boundary (just below the lower spine). Notably, the field lines follow the high values of $|\textbf{J}|/|\textbf{B}|$ on the bottom boundary and remain anchored. Panels (e) to (h) show the evolution of the same lower spine field lines during the MHD simulation. The large values of $|\textbf{J}|/|\textbf{B}|$ do not appear below the lower spine (on the bottom boundary) and it does not remain anchored from $t\approx55$ onward (panels (f) to (h)).}
\label{R2comp-CS}
\end{figure}

To compare the simulated magnetic field lines dynamics in region
R2 with the observed tip of the W-shaped flare ribbon 
B2 (Figure \ref{extrapolation} (a)) during the HMHD and MHD evolution,
we present the instants from both
simulations at t=70 in panels (a) and (b) of Figure \ref{R2comp}
respectively. Importantly, the lower spine remains anchored to the bottom boundary during the HMHD simulation (evident from the supplemented animation along with Figure \ref{R2comp}). Further, Figure \ref{R2comp-CS} shows the evolution of the lower spine along with the $|\textbf{J}|/|\textbf{B}|$ on the bottom boundary for the HMHD (panels (a) to (d)) and MHD (panels (e) to (h)) cases. In the HMHD case, noteworthy is the slipping motion of lower spine  (marked by the black arrows) tracing the $|\textbf{J}|/|\textbf{B}|>350$ regions on the bottom boundary (panels (a) to (b)). Whereas, in the MHD such high values of $|\textbf{J}|/|\textbf{B}|$ are absent on the bottom boundary---suggesting the slippage of the field lines on the bottom boundary to be less effective in contrast to the HMHD. The finding is in agreement with the idea of enhanced slippage of field lines due to high current densities as conceptualized in the introduction.   
The anchored lower spine provides a path for the plasma to flow downward 
to the brightening segment B2. In the actual corona, such flows result in 
flare brightening \citep{Benz}.
In contrast, the lower
spine gets completely disconnected from the bottom boundary (Figure
\ref{R2comp} (b)) in the MHD simulation, hence failing to explain
the tip of the W-shaped flare ribbon in B2. The
anchored lower spine in the HMHD simulation is caused by a complex
series of magnetic field lines reconnections at the 3D null and
along the QSLs in R2, as depicted in the animation. 

\subsubsection{Region R3} 
\begin{figure}
\centering
\includegraphics[width=1\linewidth]{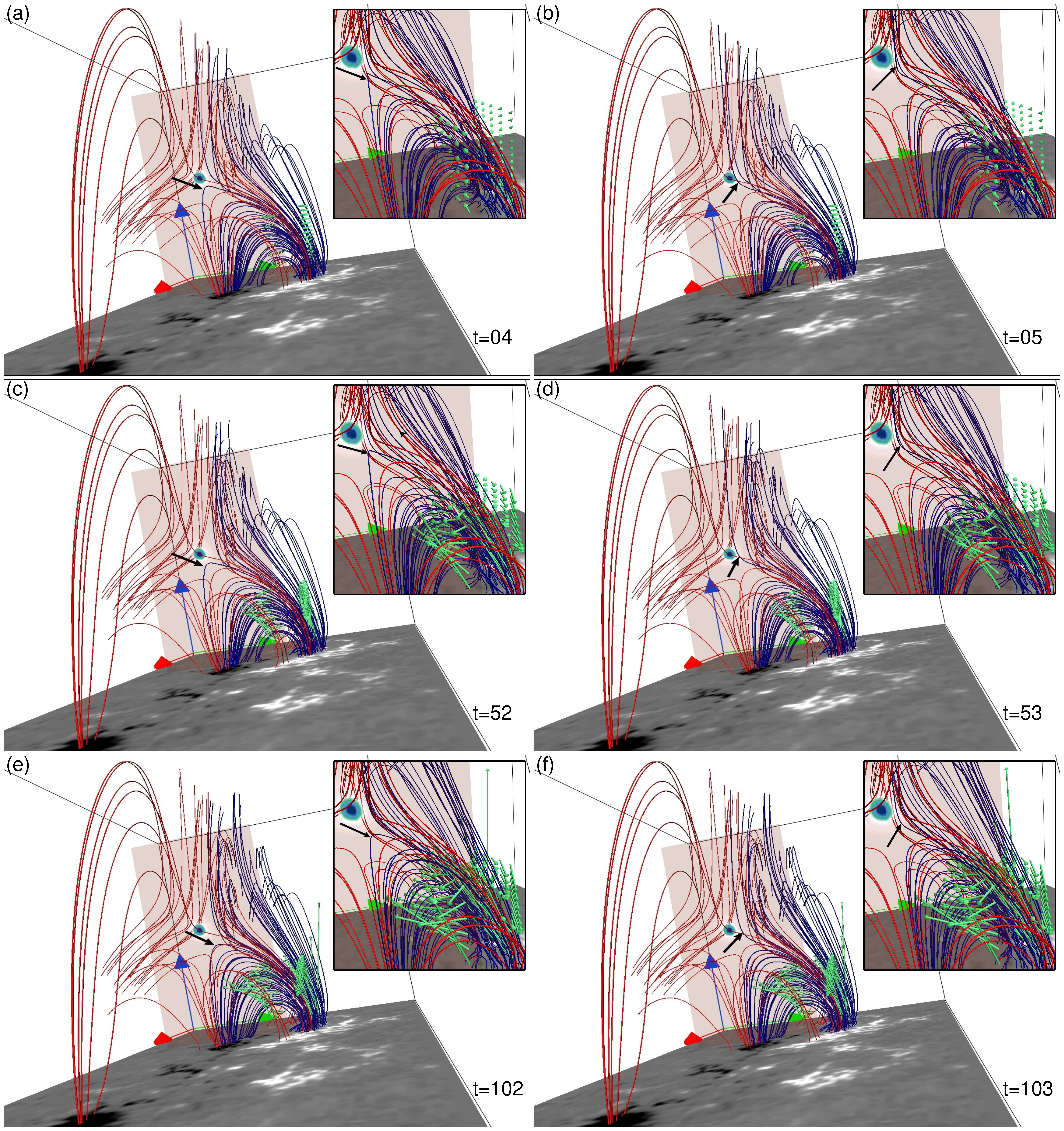}
\caption{Time sequence showing the blue field line prior and after reconnection (indicated by back arrow) in region R3 during the HMHD simulation. Evolution of the flow vectors is depicted by green arrows (on the right side)---mimicking the direction of the plasma flow. The plane along the cross section of magnetic field lines morphology in R3, showing the blue circular contours represent the value of $n$ (also shown in Figure \ref{R2R3R4exp}(d)).}
\label{R3HMHD}
\end{figure}

\begin{figure}
\centering
\includegraphics[width=1\linewidth]{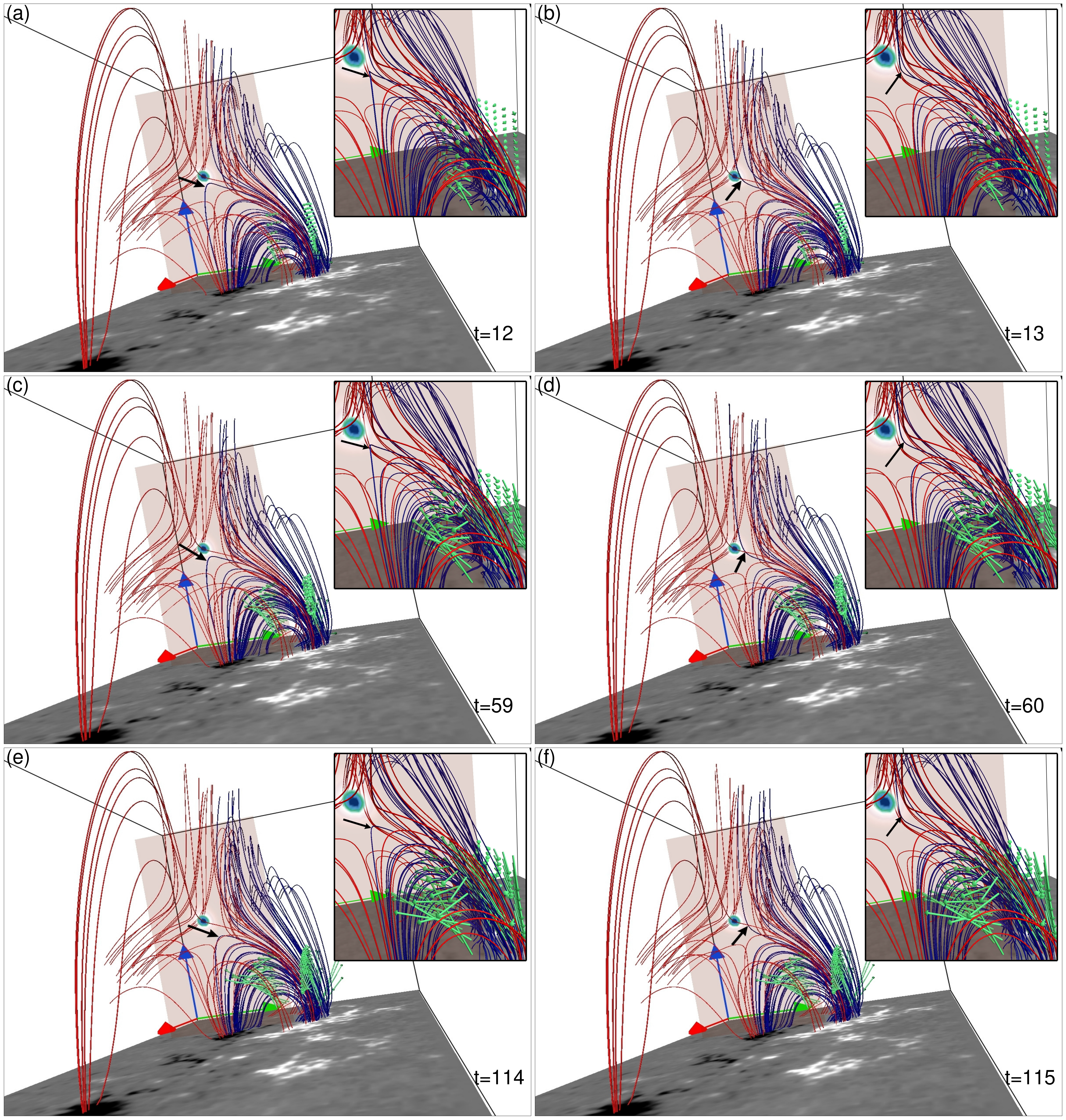}
\caption{Time sequence showing the blue field line prior and after reconnection (indicated by back arrrow) in region R3 during the MHD simulation. Evolution of the flow vectors is depicted by green arrows (on the right side)---mimicking the direction of the plasma flow. The plane along the cross section of magnetic field lines morphology in R3, showing the blue circular contours represent the value of $n$ (also shown in Figure \ref{R2R3R4exp}(d)). Notably, reconnection of the blue magnetic field lines is slightly delayed in comparison to its HMHD counterpart. A combined animation of Figure \ref{R3HMHD} and this figure is available. The video shows the evolution of magnetic field lines of region R3 along with the flow vectors (green) from t=0 to 120 for the HMHD and MHD simulations respectively. The realtime duration of the video is 12 s.}
\label{R3MHD}
\end{figure}

HMHD and MHD simulations of magnetic field lines dynamics around
the null-line are shown in Figures~\ref{R3HMHD} and \ref{R3MHD}
respectively. Figure~\ref{R3HMHD} shows the blue magnetic field
lines prior and after the reconnections (indicated
by black arrows) between t=4 to 5 (panels (a)-(b)), t=52 to 53
(panels (c)-(d)), and t=102 to 103 (panels (e)-(f)) during the HMHD
simulation. Figure \ref{R3MHD} shows the same blue
magnetic field lines prior and after the reconnections
(indicated by black arrows) between t=12 to 13 (panels (a)-(b)),
t=59 to 60 (panels (c)-(d)), and t=114 to 115 (panels (e)-(f))
during the MHD simulation.  Comparison of the panels (a)-(f) of
Figure \ref{R3HMHD} with the same panels of Figure \ref{R3MHD}
reveals earlier reconnections of the blue magnetic
field lines in the HMHD simulation. In both figures, green
velocity vectors on the right represent the local plasma flow. They
get aligned downward along the foot points of the fan magnetic field
lines, as reconnection progresses.  Consequently, the plasma flows
downward and impacts the denser and cooler chromosphere to give
rise to the brightening in B3. The velocity vectors
pointing upward represent a flow toward the null-line. The plasma
flow pattern in R3 is the same in the HMHD and in 
the MHD simulation. The vertical $yz-$plane passing through the cross section
of the null-line surface (also shown in Figure \ref{R2R3R4exp} (d))
in all the panels of Figures \ref{R3HMHD} and \ref{R3MHD} shows the
variation of $n$ with time.  It is evident that the null is not
destroyed throughout the HMHD and MHD evolution. Structural changes in the field lines caused by reconnection is near-identical for both the simulations, indicating inefficacy of the Hall term. This inefficacy is justifiable as $|\textbf{J}|/|\textbf{B}|$ remains small $\approx 10$ (not shown) in both HMHD and MHD evolution.  

\subsubsection{Region R4} 
\begin{figure}
\includegraphics[width=1\linewidth]{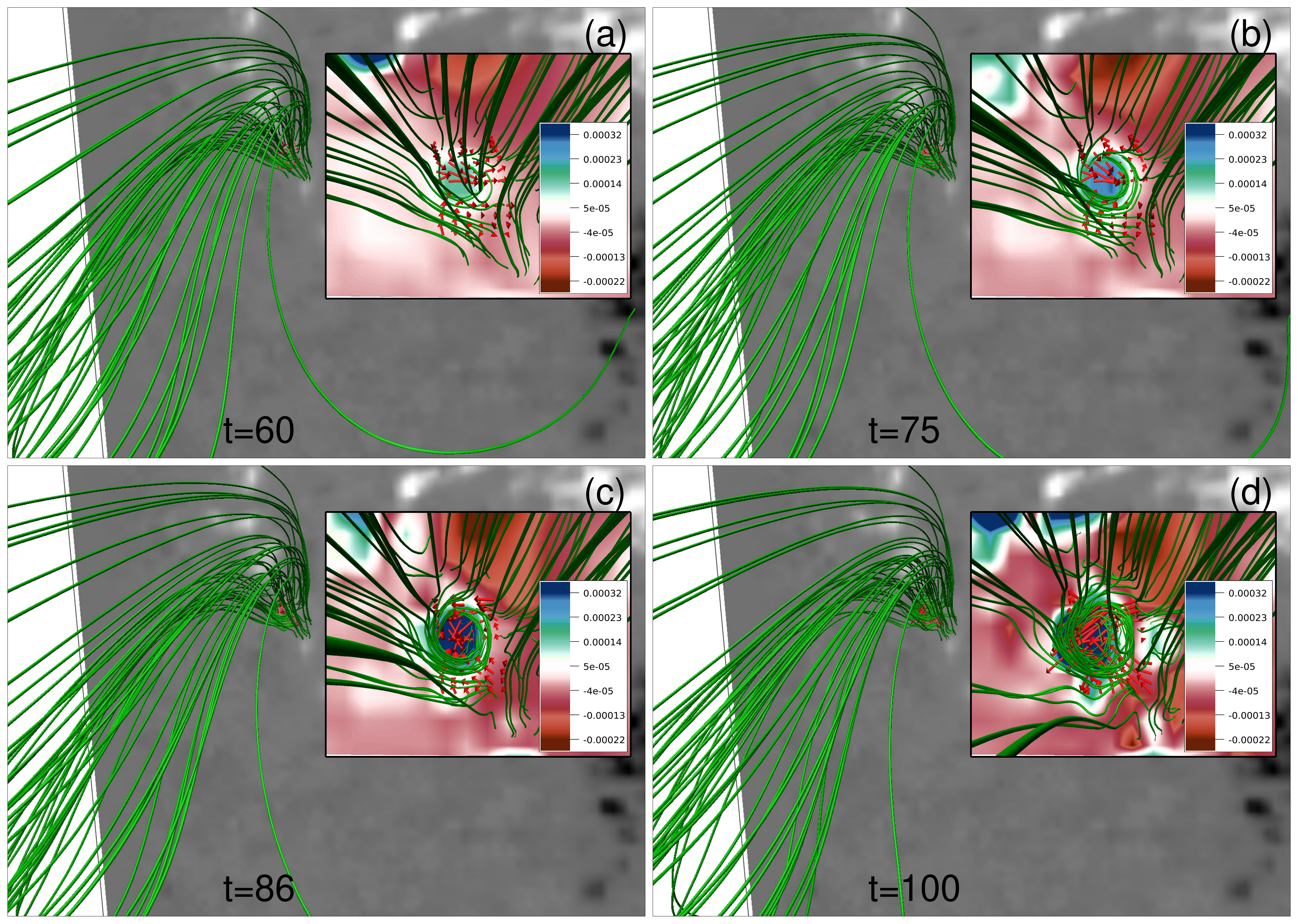}
\caption{Panels (a)-(d) show the global dynamics of magnetic field lines in region R4 during the HMHD simulation. Inset images in each panel (on right) depict the time sequence of the zoomed top-down view of the rotational motion of magnetic field lines. The background shows the variation of the z-component of flow $\in$ [-0.00022,0.00032] in all inset images. The red vectors represent the plasma flow and change its direction in an anticlockwise manner in panels (a)-(d). The rotational motion of magnetic field lines coincides with the circular part of the flow. An animation of this figure is available. The video shows the evolution of magnetic field lines in region R4 from t=0 to 100 for the HMHD simulation. The development of an anticlockwise rotational motion of the foot points from t$\approx$33 onward and the clockwise slippage of field line from t$\approx$76 onward is evident in the video. The realtime duration of the video is 10 s.}
\label{lftcrclrmotion}
\end{figure}

\begin{figure}
\includegraphics[width=1\linewidth]{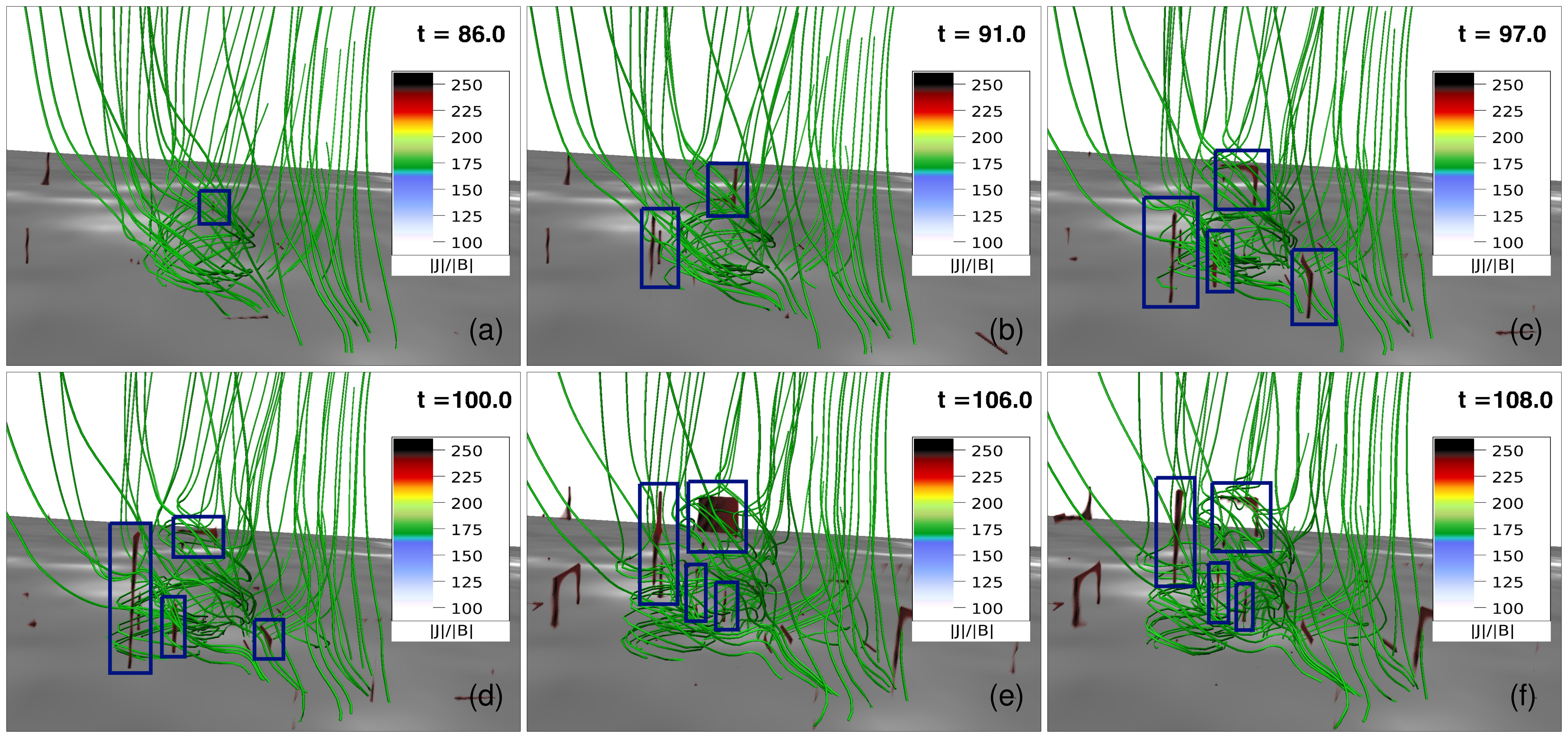}
\caption{Panels (a) to (f) show the side view of the rotating magnetic field lines structure in the region R4 overlaid with $|\textbf{J}|/|\textbf{B}|$. The figure depicts the temporal development of strong magnetic field gradient regions of $|\textbf{J}|/|\textbf{B}|>225$ (enclosed in the blue rectangular boxes) within the rotating magnetic structure.}
\label{lftcrclrmotion-SV}
\end{figure}

The development of the circular motion of magnetic field lines in region R4 during the HMHD simulation is depicted in Figure \ref{lftcrclrmotion}. It shows the global dynamics of magnetic field lines in R4 and the inset images show the zoomed view of magnetic field lines in R4 to highlight the circular motion of magnetic field lines. The bottom boundary is $B_z$ in the main figure while the inset images have the $z-$component of the plasma flow at the bottom boundary (on $xy-$plane). The red vectors represent the plasma flow direction as well as magnitude in all the panels of Figure \ref{lftcrclrmotion} where the anticlockwise pattern of the plasma flow is evident. The global dynamics highlight reconnection of the loop anchored between positive and negative  polarities at t=60 in Figure \ref{lftcrclrmotion} as it gets disconnected from the bottom boundary in panels (c)-(d) of Figure \ref{lftcrclrmotion}. The animation accompanying Figure \ref{lftcrclrmotion} highlights an anticlockwise motion of foot points in the 
same direction as the plasma flow, indicating field lines to be frozen in the fluid. 
The trapped plasma may cause the rotating structure B4 in the observations (c.f. Figure \ref{extrapolation} (a)). However, no such motion is present during the MHD evolution of the same magnetic field lines (not shown). An interesting feature noted in the animation is the clockwise slippage of field lines after the initial anticlockwise rotation. Further analysis of R4 using the direct volume rendering of $|\textbf{J}|/|\textbf{B}|$ is presented in Figure \ref{lftcrclrmotion-SV}. The figure shows $|\textbf{J}|/|\textbf{B}|$ attains high values $\ge225$ (enclosed by the blue rectangles) within the rotating field lines from t$\approx$86 onward. This suggests the slippage of field lines is, once again, related to the high magnetic field gradients.

\par
\begin{figure}
\centering
\includegraphics[width=0.55\textwidth, height=1\textwidth]{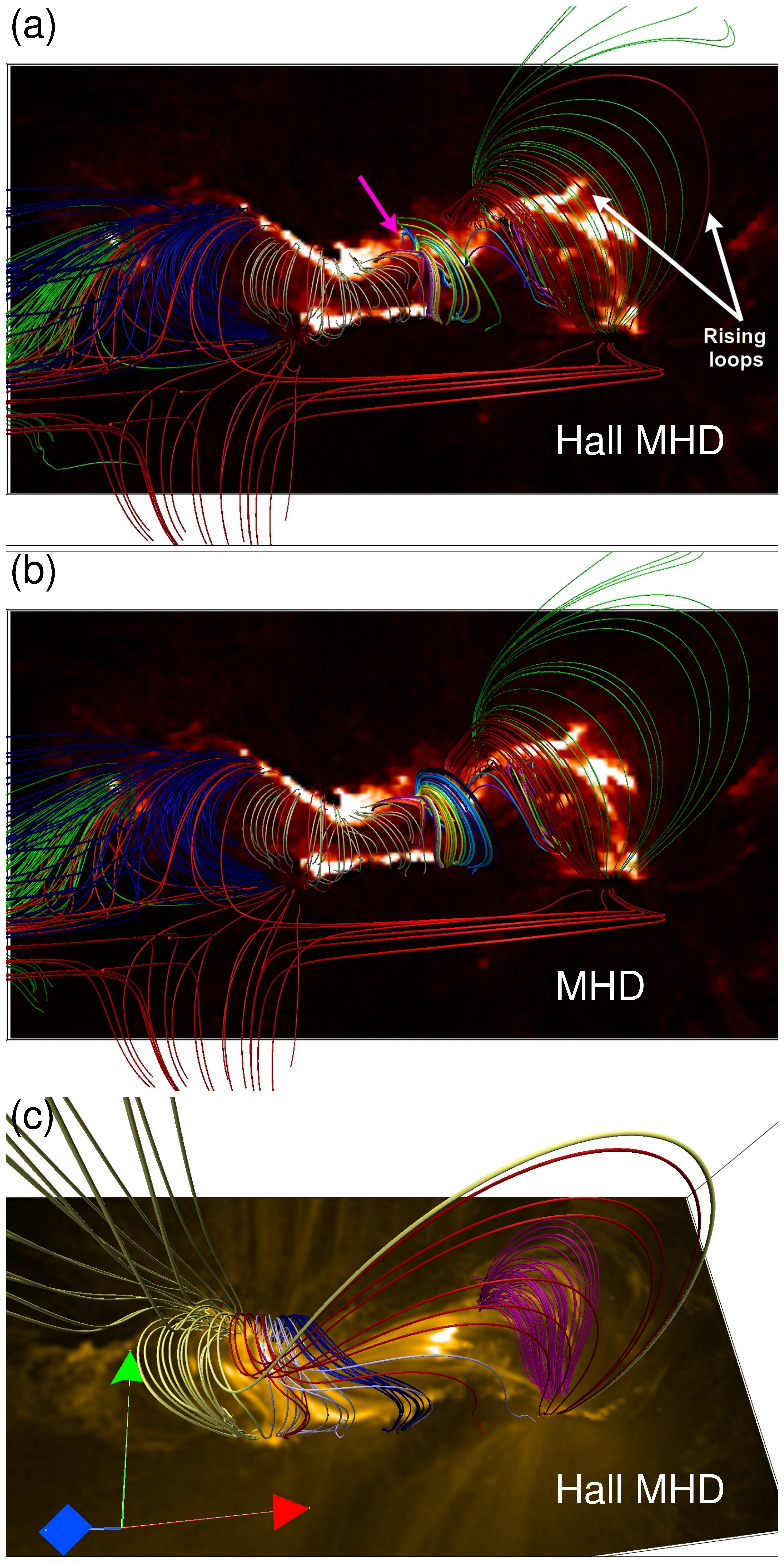}
\caption{Top-down view of an overall magnetic field lines morphology overlaid on the SDO/AIA 304 {\AA} (panels (a) and (b)) and 171 {\AA} images (panel (c)). Anchored magnetic field lines foot points in central part match well with the observed tip of the W-shaped flare ribbon (marked by pink arrow in panel (a)) in the HMHD while magnetic field lines foot points are completely disconnected from the bottom boundary in the MHD (panel (b)). Loops rising higher up in the corona is remarkable in the HMHD (indicated by white arrow in panel (a)). An animation of panels (a) and (b) is available. The video shows the evolution of magnetic field lines in AR along with the AIA 304 {\AA} image at the bottom from t=0 to 120 for the HMHD and MHD simulations respectively. The realtime duration of the video is 12 s.}
\label{Tv304171}
\end{figure}

 For completeness, we present the snapshots of an overall magnetic field lines morphology including the magnetic structures and topology of regions R1, R2, R3, and R4 together, overlaid with 304 {\AA} and 171 {\AA} from the HMHD and MHD simulations. Figure \ref{Tv304171} (a) shows an instant (at t=75) from the HMHD simulation where the topologies and magnetic structures in R1, R2, R3, and R4, plus the additionally drawn locust color magnetic field lines between R2 and R3 are shown collectively. It shows an excellent match of the magnetic field lines in R2 with the observed tip of W-shaped flare ribbon at B2, which is pointed out by the pink arrow in panel (a). Foot points of the spine-fan geometry around the 3D null orient themselves in the same fashion as the observed tip of the W-shaped flare ribbon at B2 as seen in 304 {\AA} channel of SDO/AIA. The rising loops indicated by the white arrow correspond to the same evolution as shown in Figure \ref{fullR1}. An overall magnetic field lines morphology mentioned in Figure \ref{lftcrclrmotion} (a) is given at the same time (t=75) during the MHD simulation overlaid with 304 {\AA} image in Figure \ref{lftcrclrmotion} (b). Importantly, unlike the HMHD simulation, the MHD simulation does not account for the anchored lower spine and fan magnetic field lines of the 3D null at the center of the B2. Also, the significant rise of overlying maroon magnetic field lines and the circular motion of the material in B4 is captured in the HMHD simulation only. In panel (c) magnetic field lines overlaid with 171 {\AA} image shows the magnetic field lines (higher up in the solar atmosphere) have resemblance with the post-flare loops during the HMHD. Overall, the HMHD evolution seems to be in better agreement with the observations in comparison to the MHD evolution.

\section{Summary and Discussion}
\label{summary}
The paper compares data-based HMHD and MHD simulations using the flaring Active Region NOAA 12734 as a test bed.
Importance of the HMHD stems from the realization that the Hall term in the induction equation cannot be neglected in presence of the magnetic reconnection---the underlying cause of solar flares. 
The selected event is the C1.3 class flare on March 08, 2019 around 03:19 UT for the aforementioned comparison. Although the event is analyzed and reported in the literature, it is further explored using the multi-wavelength observations from SDO/AIA. The identified important features are:
an elongated extreme ultraviolet (EUV) counterpart of the eruption on the  western  side of  the  AR,  a  W-shaped  flare  ribbon  and  circular  motion  of  cool  chromospheric material on the eastern part.
The magnetic field line dynamics near these features are utilized to compare the simulations. 
Notably, the simulations
idealize the corona to have an Alfv\`en speed which is two orders of 
magnitude smaller than its
typical value. Congruent to the general understanding, the Hall parameter is selected to tie the Hall dynamics to the dissipation scale $\mathcal{O} (\Delta \textbf{x})$
in the spirit of the ILES carried out in the paper. The magnetic reconnection here is 
associated with the slippage of magnetic field lines from the plasma parcels, effective at the dissipation scale due to local enhancement of magnetic field gradient. The same enhancement also amplifies the Hall contribution, 
presumably enhancing the slippage and thereby making the reconnection faster and more effective than the MHD.

The coronal magnetic field is constructed by extrapolating the photospheric vector magnetic field obtained from the SDO/HMI observations employing the non-FFF technique \citep{Hu2010}. The concentrated distribution of the Lorentz force on the bottom boundary and its decrease with the height justify the use of non-FFF extrapolation for the solar corona. The initial non-zero Lorentz force is also crucial in generating self-consistent flows that initiate the dynamics and cause the magnetic reconnections. 
Analyses of the extrapolated magnetic field reveal several magnetic structures and topologies of interest: a flux rope on the western part at flaring location, a 3D null point along with the fan-spine configuration at the centre, a ``Fish-bone-like structure" surrounding the null-line on the eastern part of the AR. All of these structures are found to be co-spatial with the observed flare ribbon brightening.

\par The HMHD simulation shows faster slipping reconnection of the flux rope foot points and overlying magnetic field lines (constituting QSLs above the flux rope) at the flaring location. Consequently, the overlying magnetic field lines rise, eventually reaching higher up in the corona and reconnecting to provide a path for plasma to eject out.  The finding is in agreement with the observed elongated EUV counterpart of the eruption on western part of the AR. Contrarily, such significant rise of the flux rope and overlying field lines to subsequently reconnect higher up in the corona is absent in the MHD simulation---signifying the reconnection to be slower compared to the HMHD. Intriguingly, rise and expansion of the flux rope and overlying field lines owing to slipping reconnection on QSLs has also been modelled and observed in an earlier work by \citet{dudik}. 
 These are typical features of the ``standard solar flare model in 3D'', which allows for a consistent explanation of events which are
 not causally connected \citep{dudik}. It also advocates that null-points and true separatrices are not required for the eruptive flares to occur---concurring the results of this work.
HMHD evolution of the fan-spine configuration surrounding the 3D null point is in better agreement with the tip of W-shaped flare ribbon at the centre of the AR. The lower spine and fan magnetic field lines remain anchored to the bottom boundary throughout the evolution which can account for the plasma flowing downward after the reconnection and cause the brightening. Whereas in the MHD, the lower spine gets disconnected and cannot 
account for the brightening.  The reconnection dynamics around the null-line and the corresponding plasma flow direction is same in the HMHD as well as the MHD simulation and agrees with the observed brightening. Nevertheless, reconnection is earlier in the HMHD.  HMHD evolution captures an anti-clockwise circular motion of magnetic field lines in the left part of the AR which is co-spatial with the location of the rotating chromospheric material in eastern side of the AR. No such motion was found in the MHD simulation. Importantly, the simulations explicitly associate
generation of large magnetic field gradients to HMHD compared to MHD, resulting in faster and more efficient  field line slippage because of the enhanced Hall term.

Overall, the results documented in the paper show the HMHD explains the flare brightening better than the MHD, prioritizing the requirement to include HMHD in future state-of-the-art data-based numerical simulations. 

\section{Acknowledgement} 
The simulations are performed using the 100TF cluster Vikram-100 at Physical Research Laboratory, India. We wish to acknowledge the visualization software VAPOR (\url{www.vapor.ucar.edu}), for generating relevant graphics. Q.H. and A.P. acknowledge partial support of NASA grants 80NSSC17K0016, 80NSSC21K1671, LWS 80NSSC21K0003 and NSF awards AGS-1650854 and AGS-1954503. This research was also supported by the Research Council of Norway through its Centres of Excellence scheme, project number 262622, as well as through the Synergy Grant number 810218 (ERC-2018-SyG) of the European Research Council.

\appendix
\section{}
\label{appnd}
The dimensionless time step is obtained by employing the Hall induction equation

\begin{equation}
\label{h1}
    \frac{\partial \textbf{B}}{\partial t}=-d_H{\nabla} \times (({\nabla}\times\textbf{B})\times \textbf{B}),
\end{equation}
for a stationary fluid. The aforementioned equation is linearized over an 
equilibrium magnetic field $\textbf{B}_0$ to obtain

\begin{equation}
\label{h2}
 \frac{\partial \delta \textbf{B}}{\partial t}=-d_H[\nabla\times(\nabla \times \delta \textbf{B})\times \textbf{B}_0],
\end{equation}

\noindent $\delta\textbf{B}$ being the perturbation. To obtain the wave modes, the perturbation is assumed 
to be periodic along $x$ and $y$ of a Cartesian coordinate system 

\begin{equation}
\label{h3}    
\delta B_x = \delta B_y \propto \exp[i(k_{z}z-\omega t)].
\end{equation}
\noindent where the equilibrium field is
selected as $\textbf{B}_0=B_0 \hat{e}_z$. Straightforward mathematical manipulations yield the dispersion relation for the Whistler wave as

\begin{equation}
{\omega =d_H B_0 {k_z}^2}.
\label{Dimlssdisp}
\end{equation}

\noindent The wave number is selected as $k_z = (2\pi)/{\Delta z}$, $\Delta z$ is the dissipation scale in the computational domain, making the choice harmonious with the philosophy used extensively in the paper. 
since the dimensionless $\rho_{0}=1$ in the numerical model, (\ref{Dimlssdisp}) can be written

\begin{equation}
    \left(\frac{\Delta z}{\Delta t}\right)_{whis} = 4\pi^{\frac{3}{2}} d_{H}\left(\frac{\Delta z}{\Delta t}\right)_{Alf}{\left(\frac{1}{\Delta z}\right)_{whis}}.
\end{equation}

\noindent With  $\Delta z_{whis}=\Delta z_{Alfven}=0.0078$,  namely the dissipation scale 
in the present model along with $d_H$=0.004, while $\Delta t_{Alf}\approx 10^{-3}$ from previous numerical experiments; in the present model $\Delta t= \Delta t_{whis}\approx 10^{-4}$.


\end{document}